\documentclass[twocolumn,superscriptaddress,aps,prl,longbibliography,showpacs,preprintnumbers,amsmath,amssymb,floatfix]{revtex4-1}
\usepackage{titlesec}
\usepackage{amsmath}
\usepackage{graphicx}
\usepackage[ansinew]{inputenc}
\usepackage{array}
\usepackage{color}
\usepackage{amsxtra}
\usepackage{amstext}
\usepackage{amssymb}
\usepackage{latexsym}
\usepackage{gensymb}
\usepackage{dsfont}
\usepackage{braket}
\usepackage[makeroom]{cancel}

\usepackage{subfigure}
\usepackage{dcolumn}
\usepackage{bm}
\usepackage{bbm}
\usepackage{braket}
\usepackage{mathrsfs}
\usepackage{amstext}
\usepackage{euscript}
\usepackage{txfonts}
\usepackage{footmisc}

\usepackage[unicode=true,
 bookmarks=false,
 breaklinks=false,pdfborder={0 0 1},backref=false,colorlinks=true,citecolor=blue]
 {hyperref}

\newcommand{\dg}{^\dagger}

\makeatletter
\@ifundefined{textcolor}{}
{%
 \definecolor{BLACK}{gray}{0}
 \definecolor{WHITE}{gray}{1}
 \definecolor{RED}{rgb}{1,0,0}
 \definecolor{GREEN}{rgb}{0,1,0}
 \definecolor{BLUE}{rgb}{0,0,1}
 \definecolor{CYAN}{cmyk}{1,0,0,0}
 \definecolor{MAGENTA}{cmyk}{0,1,0,0}
 \definecolor{YELLOW}{cmyk}{0,0,1,0}
}

\makeatletter
\renewcommand*\env@matrix[1][*\c@MaxMatrixCols c]{%
  \hskip -\arraycolsep
  \let\@ifnextchar\new@ifnextchar
  \array{#1}}
\makeatother

\newcommand{\eref}[1]{Eq.\,\eqref{#1}}
\newcommand{\fref}[1]{Fig.\,\ref{#1}}

\newcommand{\cref}[1]{Ref.\,\cite{#1}}

\newcommand{\mc}[1]{\mathcal{#1}}

\makeatother

\begin{document}


\title{Trapped Ion Quantum Information Processing with Squeezed Phonons}

\author{Wenchao Ge}
\affiliation{United States Army Research Laboratory, Adelphi, Maryland 20783, USA}
\affiliation{The Institute for Research in Electronics and Applied Physics (IREAP), College Park, Maryland 20740, USA}

\author{Brian C. Sawyer}
\affiliation{Georgia Tech Research Institute, Atlanta, Georgia 30332, USA}

\author{Joseph W. Britton}
\affiliation{United States Army Research Laboratory, Adelphi, Maryland 20783, USA}

\author{Kurt Jacobs}
\affiliation{United States Army Research Laboratory, Adelphi, Maryland 20783, USA}
\affiliation{Department of Physics, University of Massachusetts at Boston, Boston, Massachusetts 02125, USA}
\affiliation{Hearne Institute for Theoretical Physics, Louisiana State University, Baton Rouge, Louisiana 70803, USA}

\author{John J. Bollinger}
\affiliation{National Institute of Standards and Technology, Boulder, Colorado 80305, USA,}

\author{Michael Foss-Feig}
\affiliation{United States Army Research Laboratory, Adelphi, Maryland 20783, USA}
\affiliation{Joint Quantum Institute, NIST/University of Maryland, College Park, Maryland 20742, USA}
\affiliation{Joint Center for Quantum Information and Computer Science, NIST/University of Maryland, College Park, Maryland 20742, USA}

\begin{abstract}
Trapped ions offer a pristine platform for quantum computation and simulation, but improving their coherence remains a crucial challenge. Here, we propose and analyze a new strategy to enhance the coherent interactions in trapped ion systems via parametric amplification of the ions' motion---by squeezing the collective motional modes (phonons), the spin-spin interactions they mediate can be significantly enhanced.  We illustrate the power of this approach by showing how it can enhance collective spin states useful for quantum metrology, and how it can improve the speed and fidelity of two-qubit gates in multi-ion systems, important ingredients for scalable trapped ion quantum computation.  Our results are also directly relevant to numerous other physical platforms in which spin interactions are mediated by bosons.
\end{abstract}

\maketitle

Trapped ions are among the best developed implementations of numerous quantum technologies, including quantum computers \cite{PhysRevLett.74.4091}, quantum simulators \cite{Blatt:2012aa}, and quantum measurement devices \cite{PhysRevA.46.R6797}. For example, universal quantum gate sets have been implemented with extremely high fidelity in small systems \cite{Ballance:2016, Gaebler:2016aa}, while quantum spin dynamics and entanglement generation have been demonstrated among tens \cite{Zhang_2017} and even hundreds \cite{Bohnet2016} of ions.  For all of these applications, the general approach is to identify a qubit, i.e.,\ two metastable atomic states, and then engineer interactions between qubits by controllably coupling them to the ions' collective motion (phonons), typically using lasers \cite{PhysRevLett.74.4091,PhysRevLett.75.4714} or magnetic field gradients \cite{maggrad1, maggrad2}.  Putting aside the details of what specifically constitutes a qubit (hyperfine states of an ion, Rydberg levels of a neutral atom, charge states of a superconducting circuit), and what type of boson mediates interactions between them (phonons or photons), this basic paradigm of controllable boson-mediated interactions between qubits is at the heart of many physical implementations of quantum technologies.  In all such systems, a key technical challenge is to make the interactions as strong as possible without compromising the qubit.
\begin{figure}[t]
\leavevmode\includegraphics[width = 0.95 \columnwidth]{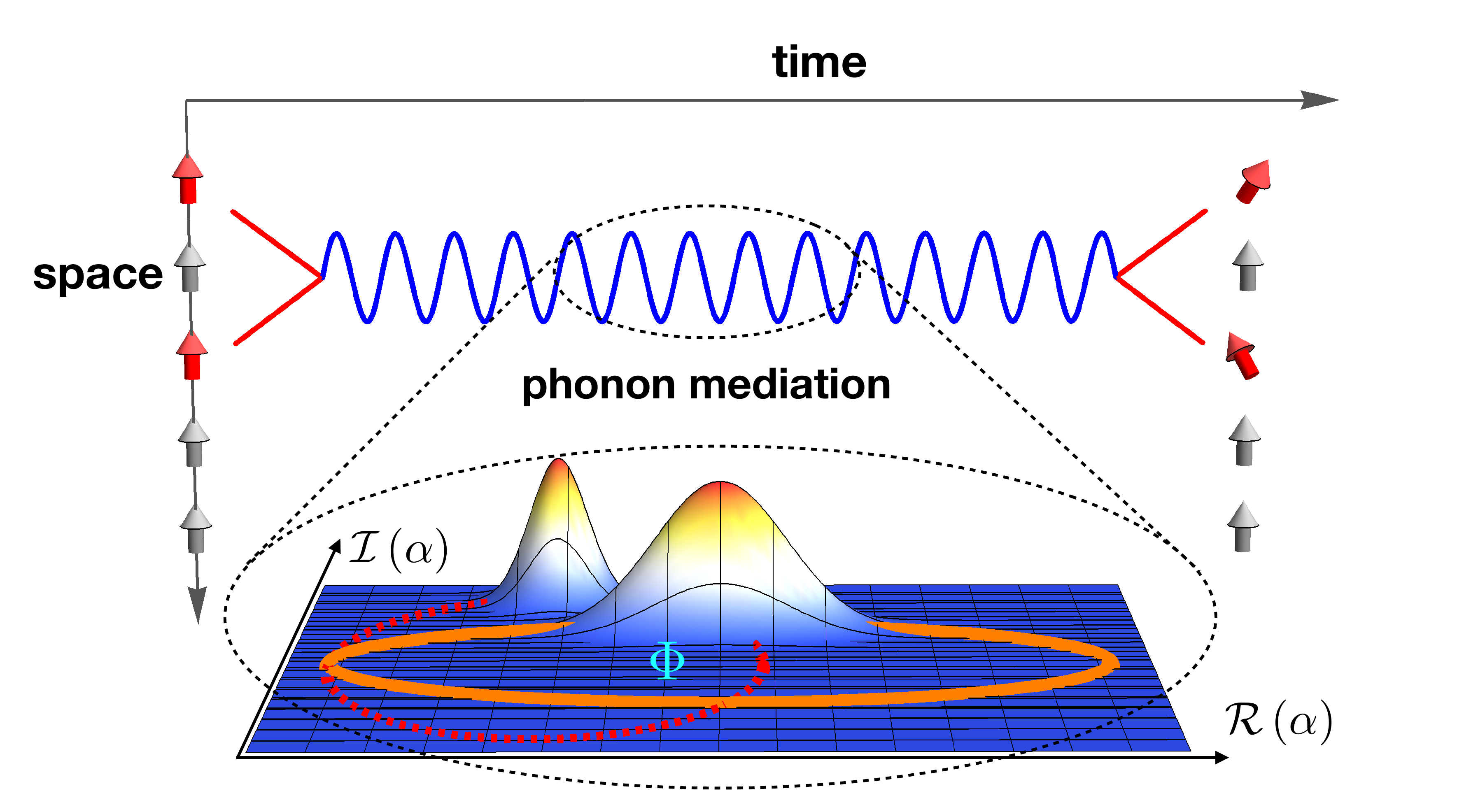}
\caption{Spin-spin interactions among trapped ions are mediated by phonon exchange, and their strength is proportional to the rate at which area ($\Phi$) is enclosed by the phonon trajectories in phase space. The trajectories enclose area faster with parametric amplification (orange ellipse) than without (red dashed circle), leading to stronger spin-spin interactions.  } 
\label{fig:traj} 
\end{figure} 

For trapped ions, the strength of interactions between qubits (from here forward called spins) is often limited by the available laser power or by the current that can be driven through a thin trap electrode.  Where these technical limitations can be overcome, other more fundamental limits remain. For example, the scattering due to the laser beams that generate spin-spin interactions can be the dominant source of decoherence \cite{Ballance:2016, Gaebler:2016aa,Bohnet2016}, in which case using more laser power is not necessarily helpful \cite{PhysRevA.75.042329,PhysRevLett.105.200401,PhysRevA.97.052301}. Moreover, in many-ion strings larger laser power can lead to decoherence through off-resonant coupling to undesirable modes, a source of decoherence that becomes more severe with increasing ion number \cite{PhysRevLett.112.190502}. (Although this effect may be mitigated, it requires modulating the laser parameters in a complicated fashion \cite{PhysRevLett.112.190502, PhysRevLett.114.120502, PhysRevLett.120.020501}.)  In this Letter, we propose a straightforward experimental strategy to increase the strength of boson-mediated spin interactions that can also overcome the aforementioned limitations, and is sufficiently flexible to be relevant to numerous other systems in which qubits interact by exchanging bosons.  In particular, we consider modulating the ions' trapping potential at nearly twice the typical motional mode frequency \cite{Heinzen1990}.  Related forms of parametric amplification (PA) of boson-mediated interactions have been considered recently in systems ranging from phonon-mediated superconductivity \cite{Babadi2017}, to optomechanics \cite{lemonde2016enhanced} and cavity or circuit QED  \cite{PhysRevX.7.021041, arenz2018hamiltonian}. Our work goes further in that we determine the effects of PA in a driven multimode system, provide a simple physical explanation of its effects based on amplified geometric phases (see \fref{fig:traj}), and determine the capability of PA to enhance specific quantum information tasks performed with trapped ions.

%
%
%

Typically, spin-spin interactions between trapped ions are induced through spin-dependent acquisition of area swept out by phonon trajectories in phase space (\fref{fig:traj}). An area $\Phi$ produces a multiplicative phase $e^{-i\Phi}$ of the corresponding spin state, a geometric phase that depends only on the enclosed area \cite{Sorensen00, Leibfried03,G-Ripoll05}. The spin dependence can be achieved by driving the ions' motion with a spin-dependent force (SDF), with characteristic interaction energy $f$ (defined below).  Once spin-dependent displacements have been seeded by the SDF, they can be amplified \emph{spin independently} by modulating the trapping potential with a carefully chosen phase relative to the applied SDF (\fref{fig:traj}). Without PA, the time it takes to accumulate a particular geometric phase $\Phi$---corresponding to the generation of a particular entangled spin state---is lower bounded by $t_{\rm min}\propto\sqrt{\Phi}/f$. With PA this scaling is modified to
\begin{align}
t_{\rm min} \propto \mathscr{S}\!\sqrt{\Phi}/f,\label{eq:PAscaling}
\end{align}
where  $\mathscr{S}<1$ is the degree of squeezing in the squeezed mechanical quadrature, enabling a particular entangled state to be created faster for fixed laser power or magnetic field gradient.



\emph{Trapped ion quantum simulators}.---Before describing the effects of PA, we briefly review the standard mechanism by which a trapped ion crystal with $N$ ions can be made to simulate the quantum Ising model \cite{Blatt:2012aa}, 
\begin{align}
\hat{\mathcal{H}}=\hbar\frac{1}{N}\sum_{i<j}J_{ij}\hat{\sigma}_{i}^{z}\hat{\sigma}_{j}^{z}
\label{eq:Isingmodel}.
\end{align}
Here, $\hat{\sigma}_{i}^{z}$ is the $z$-Pauli matrix for the $i$th ion, with the spin degree of freedom realized by two long-lived states.

In the Lamb-Dicke regime \cite{SM}, the Hamiltonian describing an SDF oscillating at frequency $\mu$ and with peak force $F$ can be written in a frame rotating at $\mu$ as \cite{PhysRevLett.101.090502, Britton2012} 
\begin{align}
\label{eq:NHodf}
\hat{\mathcal{H}}_{\rm SDF}&=\hbar \!\sum_{m=1}^{N}\Big(f_m\big(\hat{a}_m+\hat{a}_m^{\dagger}\big)\!\sum_{i=1}^N \!U_{i,m}\hat{\sigma}_{i}^{z}-\delta_m\hat{a}^{\dagger}_m\hat{a}_m\Big)+\hat{\mathcal{H}}_{\rm CR}.
\end{align}
Here, $f_m\propto F z_{0m}$ is the coupling strength of the SDF to the $m$th collective motional mode, with $z_{0m}\equiv\sqrt{\hbar/2M\omega_m}$ the characteristic length scale of that mode, $\omega_m$ its frequency, and $M$ the ion mass.  The $U_{i,m}$ are matrix elements of the normal mode transformation matrix \cite{james1998quantum}, and $\delta_m\equiv\mu-\omega_m$. The counterrotating Hamiltonian $\hat{\mathcal{H}}_{\rm CR}$ \cite{SM} can often be justifiably neglected in the rotating wave approximation (RWA).

 
There are two situations in which \eref{eq:NHodf} reduces approximately to \eref{eq:Isingmodel}.  If all of the modes are far off resonance ($\delta_m\gg f_m$), they can be eliminated adiabatically to give the effective spin-spin interaction in \eref{eq:Isingmodel} \cite{PhysRevLett.92.207901,Kim2009,SM}.  Alternatively, even if $f_m \gtrsim\delta_m$ for a single mode, as long as all other modes are far off resonance then the spin state approximately disentangles from the motional state at times that are integer multiples of $2\pi/\delta_m$. At these times the spin-state evolution is the same as that given by \eref{eq:Isingmodel}, with $J_{ij}\propto U_{i,m}U_{j,m}\times(Nf_m^2/\delta_m)$. For example, if $\mu$ is detuned close to the center of mass (COM) mode ($m=1$), then $J_{ij}=J\equiv 2f_1^2/\delta_1$, describing all-to-all interactions. (In what follows, we will drop the explicit subscripts on $f$ and $\delta$ when discussing a single mode.)

To understand the dependence of the geometric phase on the system parameters, we can consider the phase $\Phi$ acquired by a single spin for simplicity.  There is some freedom in how $\Phi$ is generated, namely, the phonon trajectory can undergo any integer number of loops, each contributing $4\pi(f/\delta)^2$ to $\Phi$ and taking a time $2\pi/\delta$. At fixed $f$, reducing $\delta$ decreases the time $t$ required to generate $\Phi$, but $\delta$ can only be reduced to the point where $\Phi=4\pi(f/\delta)^2$ because at least one loop must close. At this point, $\delta_{\rm min}=f/\sqrt{\Phi/4\pi}$, giving $t_{\rm min}=2\pi/\delta_{\rm min}\propto\sqrt{\Phi}/f$ as asserted above \eref{eq:PAscaling}.  In experiments that employ optical dipole forces to generate the SDF, the dominant decoherence source can be scattering from the laser beams that occurs at a rate $\Gamma\propto f$  \cite{PhysRevLett.105.200401,Britton2012}. In such cases, preparation of a particular entangled spin state (corresponding to a particular $\Phi$) is accompanied by the minimal accumulated decoherence $\Gamma t_{\rm min}\propto \sqrt{\Phi}$.



\emph{Parametric amplification}.---We now consider what happens when the ion motion is parametrically amplified while simultaneously being driven by the SDF.
If the PA is at twice the SDF frequency, then in a frame rotating at $\mu$ the PA Hamiltonian is \cite{Heinzen1990, SM}
\begin{align}
\hat{\mathcal{H}}_{\rm PA}&=\sum_{m}\hbar g_m\cos\left(2\mu t-\theta\right)\left(\hat{a}_me^{i\mu t}+\hat{a}_m^{\dagger}e^{-i\mu t}\right)^2.
\end{align}
Here, $g_m=eV/(M\omega_md_T^2)$, with $V$ the parametric drive voltage amplitude and $d_T$ a characteristic trap dimension. Typically $g_m$ depends weakly on $m$, and for simplicity we ignore the $m$ dependence in what follows. Values of $g$ as large as $0.1\times\omega_1$ appear feasible, in particular for traps with small $d_T$. The relative phase $\theta$ between the PA and SDF can in principle be chosen at will. We assume $\theta=0$, which is optimal; limitations imposed by fluctuations of $\theta$ have been carefully analyzed and are discussed later.

At first inspection, evolution under both $\hat{\mc{H}}_{\rm SDF}$ and $\hat{\mc{H}}_{\rm PA}$ seems complicated. $\hat{\mc{H}}_{\rm PA}$ squeezes the motional state, while $\hat{\mc{H}}_{\rm SDF}$ entangles the spin and squeezed motional states in a complicated way.  However, under the condition $0<g<\delta_m$ \footnote{In general, the condition is $|g|<|\delta_m|$. In this work, we assume $0<g<\delta_m$ for simplicity.}, each mode will still undergo a closed loop in phase space \cite{carmichael1984squeezing}, returning to the initial unsqueezed motional state and disentangling from the spin state at integer multiples of $2\pi/(\delta^{\prime}_m)$, with $\delta^{\prime}_m\equiv\sqrt{\delta_m^2-g^2}$. The total Hamiltonian can be written in a simple form by using a Bogoliubov transformation $\hat{b}_m=\cosh r_m \hat{a}_m -\sinh r_m \hat{a}_m^{\dagger}$, with $r_m=-\log\mathscr{S}_m$ and $\mathscr{S}_m=[\left(\delta_m-g\right)/\left(\delta_m+g\right)]^{1/4}$ \cite{lemonde2016enhanced}. In terms of these transformed operators, $\hat{\mathcal{H}}_{\rm T}=\hat{\mathcal{H}}_{\rm SDF}+\hat{\mc{H}}_{\rm PA}$ is given by
\begin{align}
\hat{\mathcal{H}}_{\rm T}&=\hbar\sum_{m=1}^{N}\Big( f^{\prime}_m\big(\hat{b}_m+\hat{b}^{\dagger}_m\big)\sum_{i=1}^N U_{i,m}\hat{\sigma}_{i}^{z}-\delta^{\prime}_m \hat{b}_m^{\dagger}\hat{b}_m\Big)+\hat{\mathcal{H}}_{\rm CR}, \label{eq:BmodeH}
\end{align}
where $ f^{\prime}_m= f_m/\mathscr{S}_m$ and $\hat{\mathcal{H}}_{\rm CR}$ now contains the counterrotating terms from both $\hat{\mc{H}}_{\rm SDF}$ and $\hat{\mc{H}}_{\rm PA}$ \cite{SM}. Therefore, we obtain a Hamiltonian that is identical (in the RWA) to $\hat{\mathcal{H}}_{\rm SDF}$ but with rescaled drive strengths and detunings.  Although every mode is squeezed by PA, a single mode (we assume the COM\ mode) will dominate the dynamics if $\delta-g\ll\delta_{m\neq 1}-g$. Table \ref{tab:rescaled_parameters} shows both the geometric phase $\Phi$ and duration $\tau$ of a single loop for the COM\ mode, along with the typical phase-space amplitudes $\alpha_m$ of the other modes, in the limit that $\delta-g\ll \delta+g$ (such that $\delta^{\prime}\approx 2 \delta \mathscr{S}^2$). Note that $\delta_m-g$ is bounded by the gap between the c.o.m mode and its closest neighbor, so that residual displacements $\alpha_m$ of the spectator modes are upper bounded as $1/\mathscr{S}$ increases \cite{SM}.
%

%

\renewcommand{\arraystretch}{1.4}
\setlength{\tabcolsep}{8pt}
 
\begin{table}
\caption{Rescaling of key quantities under PA.}
\begin{ruledtabular}
\begin{tabular}{c c c c}
& $\Phi$   & $ \tau$  &  $\alpha_m$ \\
\hline
SDF only & $4\pi(f/\delta)^2$ & $2\pi/\delta$ & $2f/\delta_m$\\ 
SDF+PA & $4\pi(f/\delta)^2/(4\mathscr{S}^6)$ \quad & $(2\pi/\delta)/(2\mathscr{S}^2)$ & $2f/(\delta_m-g)$
\end{tabular}
\end{ruledtabular}
\label{tab:rescaled_parameters}
\end{table}

As argued above, without PA the fastest strategy for obtaining a particular geometric phase $\Phi$ at fixed $f$ is to choose $\delta$ such that the COM\ mode undergoes a single loop, giving $t_{\rm min}\propto\sqrt{\Phi}/f$.  With PA, we can similarly argue that the optimal strategy to obtain $\Phi$ at fixed $f$ \emph{and} $\mathscr{S}$ is to choose $\delta$ such that a single loop is closed. Solving $\Phi=4\pi(f/\delta_{\rm min})^2/(4\mathscr{S}^6)$ for $\delta_{\rm min}$ [and using $t_{\rm min}=(2\pi/\delta_{\rm min})/(2\mathscr{S}^2)$] gives $t_{\rm min}\propto \mathscr{S}\!\sqrt{\Phi}/f$, as claimed in \eref{eq:PAscaling}.  Thus, we can generate the same spin state faster at fixed laser power or fixed current by reducing $\mathscr{S}$, which serves as a figure of merit for the benefits of PA.  Physically, PA squeezes the phase-space loops into ellipses (see \fref{fig:traj}), which enclose more area (per unit time) for a fixed SDF. For the important situation where the SDF is generated by optical dipole forces and the decoherence rate $\Gamma$ scales with the laser intensity, the accumulated decoherence can now be written as
\begin{align}
\Gamma t_{\rm min}\propto \mathscr{S}\!\sqrt{\Phi},
\label{eq:decoherence}
\end{align}
indicating that in principle the effect of decoherence in generating a particular entangled spin state can be made arbitrarily small.  In practice there will be limits on $ \mathscr{S}$, for example, due to the breakdown of the RWA (see \fref{fig:region}). For the illustrations that follow, all results based on the RWA have been verified by numerically solving for the dynamics of $\hat{\mc{H}}_{T}$.  In cases where the RWA is borderline, we then determine the reduction of the product $f t_{\rm min}$ for fixed $\Phi$ numerically \footnote{The geometric phase is obtained by numerically solving \eref{eq:BmodeH}, and extracting the area enclosed by the phase-space trajectories in the interaction picture of the total quadratic terms.}, and report this reduction as the effective degree of squeezing $\mathscr{S}_{\rm eff}$.

\emph{Improving quantum spin squeezing}.---As an exemplary application of PA, we show how it improves quantum spin squeezing (QSS). QSS characterizes the reduction of spin noise in a collective spin system, and is important for both entanglement detection \cite{PhysRevLett.99.250405} and precision metrology \cite{ma2011quantum}. Here, we investigate the Ramsey squeezing parameter $\xi_{R}$ \cite{PhysRevA.50.67}; for coherent spin states, $\xi_R^2=1$, while for spin squeezed states $\xi_R^2<1$\cite{ma2011quantum}.


A simple way to realize QSS is via single-axis twisting \cite{PhysRevA.47.5138}, for which the ideal minimal squeezing parameter scales as $N^{-2/3}$ for $N\gg1$ \cite{PhysRevA.47.5138, ma2011quantum}. This limit is very challenging to achieve for large $N$. In fact, for decoherence attributable to spontaneous spin flips  in the Ising ($z$) basis at a rate $\Gamma$  \cite{PhysRevLett.105.200401,foss2013nonequilibrium}, $\xi_{R}$ actually saturates for large $N$ to the asymptotic value $3[\Gamma/(2J)]^{2/3}$ \cite{SM, PhysRevLett.121.070403}, with the saturation taking place when $N\gg2J/\Gamma$. To improve spin squeezing, the ratio $J/\Gamma$ must be improved, which can be achieved via PA.  To benchmark potential improvements, we analyze the effects of PA quantitatively under the experimental conditions in Ref.\ \cite{Bohnet2016}. In \fref{fig:ff}(a), we plot the optimal spin squeezing as a function of $N$. The two outer lines represent SDF-only cases with (solid line) and without (dashed line) decoherence \footnote{For consistency with the experiment reported in Ref.\ \cite{Bohnet2016}, in the numerical simulations we include the effects of both spontaneous spin flips and additional elastic dephasing, reporting the combination of these two rates as $\Gamma$ in \fref{fig:ff}.}. The two intermediate lines show how the decoherence-free results are approached as $\mathscr{S}_{\rm eff}$ is decreased.  Figure \ref{fig:ff}(b) is similar to \fref{fig:ff}(a), but shows $\xi_{R}^2$ as a function of $1/\mathscr{S}_{\rm eff}$ for different $N$.

\begin{figure}
\centering
\includegraphics[width=1.0\columnwidth]{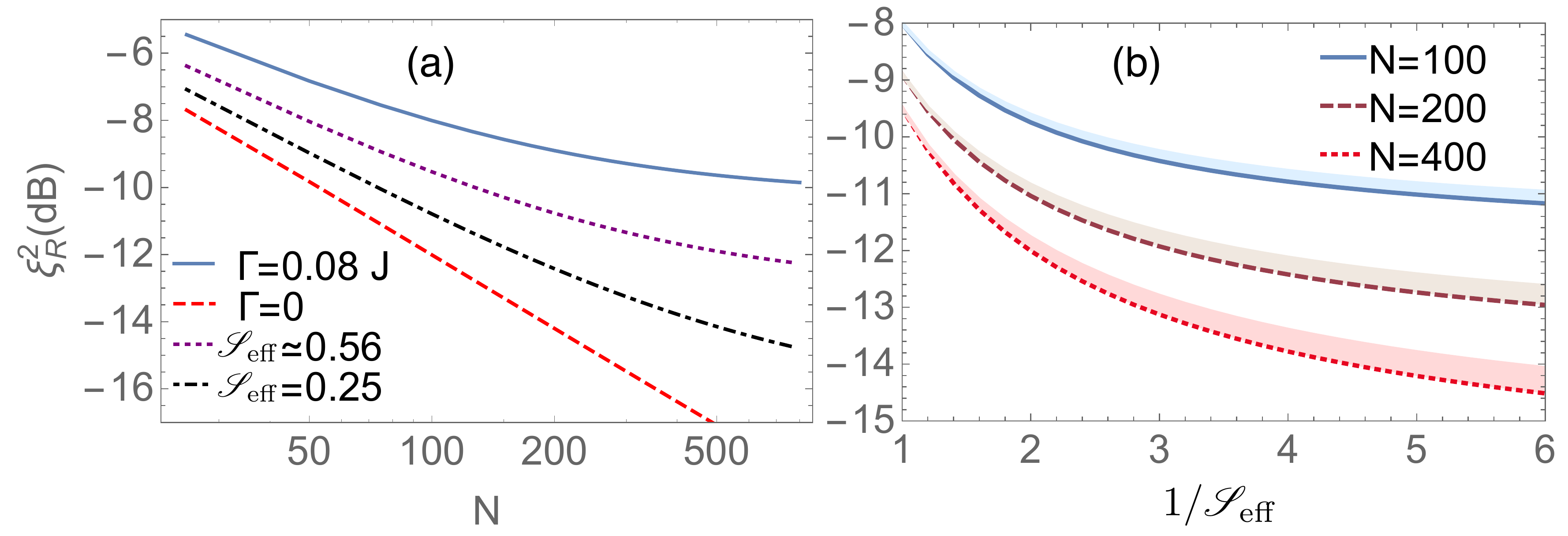}
 \caption[]{Minimal squeezing $\xi_{R}^2$ plotted (a) as a function of $N$ for various situations; (b) versus $1/\mathscr{S}_{\rm eff}$ for several values of $N$, with shaded strips indicating the expected degradation of squeezing due to a phase uncertainty of $\sigma_{\theta}=18^{0}$.}
\label{fig:ff}
\end{figure}



\emph{High fidelity two-qubit gate}.---Two-qubit gates with fidelity higher than $99.9\%$ have recently been demonstrated in two-ion systems \cite{Ballance:2016, Gaebler:2016aa}, where the largest remaining error is due to spontaneous emission from the driving lasers. Since a gate operation corresponds to some fixed $\Phi$, \eref{eq:decoherence} implies that the effective spontaneous emission rate can be reduced by a factor of $\mathscr{S}$ for a fixed gate time. 

In many-ion systems, the gate time must be much longer than the inverse of the motional mode splitting in order to suppress gate errors due to spin-phonon entanglement with off-resonant modes \cite{PhysRevLett.112.190502}. If the gate time is reduced by using more laser power, then off-resonant modes experience larger phase-space excursions  ($\alpha_m \propto f$) and the fidelity suffers. By using PA, the gate time ($\tau$) and the off-resonant loop size ($\alpha_m$) are independent, and we can hold the gate time fixed while decreasing $\alpha_m$ by a factor of $\mathscr{S}$. For example, comparing with the latest modulated pulsed laser scheme \cite{PhysRevLett.120.020501} that used $f/2\pi=10$ kHz for a two-qubit gate in a 5-ion chain, we calculate that our scheme can implement the same task with a comparable gate time ($\tau\sim180\,\mu{\rm s}$) and fidelity $\geq 99.5 \%$ using significantly less laser power (see \fref{fig:fidelity}) for the same trap frequency ($\omega_1/2\pi=3.045$ MHz). As shown in \fref{fig:fidelity}, the fidelity can be improved by tuning $g$ to minimize the total residual displacements \cite{SM}.  With the access to larger $g\sim 2\pi\times 100$ kHz, PA could enable  a much faster two-qubit gate ($\sim 30\,\mu$s) with high fidelity using moderate laser power ($f/2\pi\sim9$ kHz). \\ 

\begin{figure}[t]
\leavevmode\includegraphics[width = 0.75 \columnwidth]{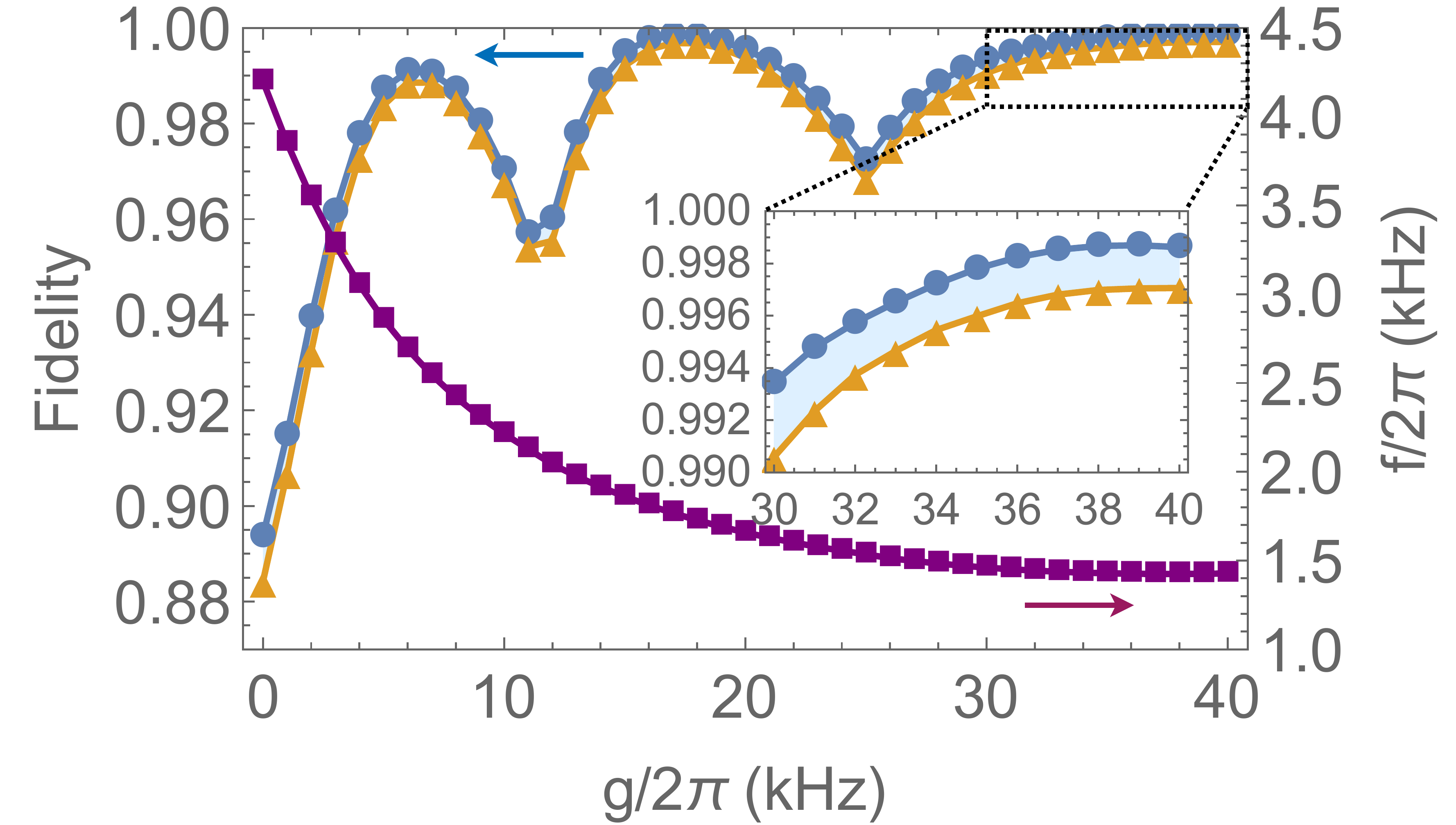}
\caption{Two-qubit gate fidelity in a $5$-ion system, calculated from a numerical simulation of the full Hamiltonian $\hat{\mathcal{H}}_{\rm T}$. The optimal fidelity with (orange triangles) and without (blue dots) timing error ($1\%$) as a function of the PA strength $g$ for a gate time $\tau\sim180\,\mu{\rm s}$. The purple squares correspond to the reduction of the laser power ($f$) as the PA strength is increased.} 
\label{fig:fidelity} 
\end{figure}

\emph{Limitations}.---Our analytical results have been simplified by dropping $\hat{\mc{H}}_{\rm CR}$ in \eref{eq:BmodeH}.  However, when the RWA breaks down the enhancement due to PA can no longer be understood simply in terms of the quadrature squeezing $\mathscr{S}$. 
Energy shifts of the Bogoliubov modes due to $\hat{\mathcal{H}}_{\rm CR}$ can be calculated in second-order perturbation theory as $\Delta\delta^{\prime}_{m}=(g_m/\mathscr{S}_m)^2/(4\mu)$, and can be ignored as long as  $\Delta\delta^{\prime}_{m}\ll \delta^{\prime}_m$ \cite{SM}, providing a necessary condition for the validity of the RWA.  To assess the validity of the RWA more quantitatively, we compare $\mathscr{S}$ with the effective degree of squeezing $\mathscr{S}_{\rm eff}$. In \fref{fig:region}, we plot both $1/\mathscr{S}^2_{\rm eff}$ and $1/\mathscr{S}^2$ as a function of $g$ for different values of the time $\tau$ for a single-loop gate with $\omega_1/2\pi=3.045$ MHz. As expected, we observe that they agree very well for small enhancement, deviating appreciably only once $\Delta\delta^{\prime}_{m}= \delta^{\prime}_m/2$ (dot-dashed region).  Note that the maximum achievable enhancement increases with increasing $\tau$. The above analysis may have implications for the limitations of PA in other systems \cite{lemonde2016enhanced}. \\

The primary technical concerns in implementing PA experimentally are likely to be the uncertainty in the relative phase $\theta$ between the SDF and the PA, shot-to-shot frequency fluctuations of $\delta$, and imperfect control of the interaction time. Controlling the phase of an optical-dipole force has been demonstrated \cite{PhysRevLett.116.033002} but can be challenging. Nonzero $\theta$ does not affect the period of a single loop, but it does reduce the geometric phase $\Phi$ enclosed by that loop, and therefore reduces the resulting spin-spin interaction strength $J$. However, we can show that $J$ depends on $\theta$ only to second order. For both spin squeezing and two-qubit gates, the figures of merit (squeezing amount and gate fidelity, respectively) scale quadratically with the shift of $J$ around its maximal ($\theta=0$) value \cite{SM}, and therefore depend only quartically on $\theta$. Modeling the phase as a zero-mean Gaussian random number with standard deviation $\sigma_{\theta}$, in \fref{fig:ff}(c) we show the expected standard deviation in $\xi^2_{\rm R}$ for $\sigma_{\theta}=18^0$. Fluctuations of $\delta$ (due to fluctuations of either $\mu$ or $\omega_m$) affect the gate fidelity quadratically by modifying the Bogoliubov frequencies $\delta^{\prime}_m$ \cite{SM}. For the simulation shown in \fref{fig:fidelity}, we estimate that fidelity $>99\%$ is still possible with shot-to-shot frequency fluctuations of $0.2$ kHz. Imperfect timing control has a similar effect as fluctuations in $\delta$ on the degree of spin squeezing and the fidelity of two-qubit gates. In \fref{fig:fidelity} we show that a $1\%$ timing error \cite{Ballance:2016}, reduces the gate fidelity by about $0.3\%$ in the $5$-ion system studied. Finally, we note that in the RWA, $\hat{\mathcal{H}}_{\rm T}$ (\eref{eq:BmodeH}) and $\hat{\mathcal{H}}_{\rm SDF}$ (\eref{eq:NHodf}) have the same form, implying that the enhancements of PA are insensitive to the temperature of the initial motional state \cite{molmer1999multiparticle} in the Lamb-Dicke regime.



\begin{figure}[t!]
\leavevmode\includegraphics[width = .7\columnwidth]{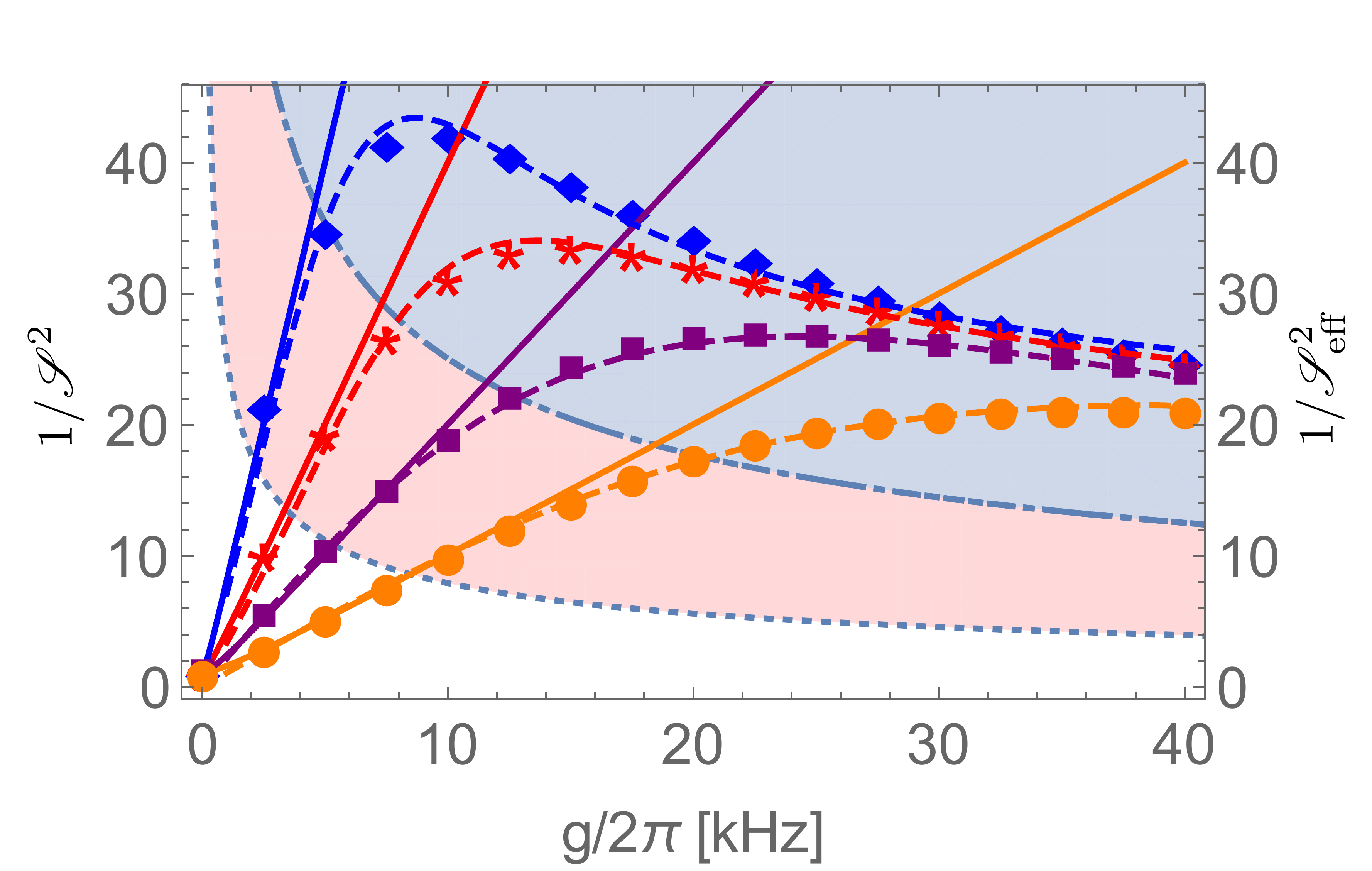}
\caption{Breakdown of the rotating wave approximation. The enhancement factors $1/\mathscr{S}^2$ (solid lines) and $1/\mathscr{S}^2_{\rm eff}$ (points, with dashed lines as guides to the eye) as a function of $g$ for different $\tau=2\pi/\delta^{\prime}\approx(2\pi/g)/(2\mathscr{S}^2)$ at $4, 2, 1, 0.5$ ms from left to right. The dotted region (red shaded) corresponds to  $\Delta \delta^{\prime}_{m}\ge \delta^{\prime}_m/20$ and the dot-dashed region (blue shaded) corresponds to $\Delta \delta^{\prime}_{m}\ge \delta^{\prime}_m/2$. } 
\label{fig:region} 
\end{figure} 

\emph{Outlook.---}
To be concrete we have focused on spin squeezing and two-qubit gates, but the techniques described here are likely to have numerous other applications.  For example, it should be possible to enhance the creation of deeply oversqueezed (non-Gaussian) spin states, and it may also be possible to improve amplitude sensing of mechanical displacements \cite{gilmore2017amplitude}.  Our strategy is not exclusive of other tools in the trapped ion toolbox; for example, it may be possible to use PA in conjunction with dynamical controls over the driving laser to further suppress unwanted spin-motion entanglement in two-qubit gates.  Similar to time-dependent control schemes \cite{PhysRevLett.112.190502, PhysRevLett.114.120502, PhysRevLett.120.020501}, we can also utilize stroboscopic parametric driving protocols to optimize the amplification of spin-spin interactions. For example, stroboscopic protocols consisting of alternating applications of a resonant SDF and a resonant PA with large $g$ can potentially increase the enhancement factor limits from the RWA breakdown.

\begin{acknowledgments}
We thank Justin Bohnet, Shaun Burd, Daniel Kienzler, and Jiehang Zhang for useful discussions. This manuscript is a contribution of
NIST and not subject to U.S. copyright. This work was supported by the Assistant Secretary of Defense for Research \& Engineering as part of the Quantum Science and Engineering Program. 
\end{acknowledgments}


\bibliography{Improving-Trapped-Ion-quantum-simulators-via-Parametric-Amplification}

\begin{thebibliography}{47}%
\makeatletter
\providecommand \@ifxundefined [1]{%
 \@ifx{#1\undefined}
}%
\providecommand \@ifnum [1]{%
 \ifnum #1\expandafter \@firstoftwo
 \else \expandafter \@secondoftwo
 \fi
}%
\providecommand \@ifx [1]{%
 \ifx #1\expandafter \@firstoftwo
 \else \expandafter \@secondoftwo
 \fi
}%
\providecommand \natexlab [1]{#1}%
\providecommand \enquote  [1]{``#1''}%
\providecommand \bibnamefont  [1]{#1}%
\providecommand \bibfnamefont [1]{#1}%
\providecommand \citenamefont [1]{#1}%
\providecommand \href@noop [0]{\@secondoftwo}%
\providecommand \href [0]{\begingroup \@sanitize@url \@href}%
\providecommand \@href[1]{\@@startlink{#1}\@@href}%
\providecommand \@@href[1]{\endgroup#1\@@endlink}%
\providecommand \@sanitize@url [0]{\catcode `\\12\catcode `\$12\catcode
  `\&12\catcode `\#12\catcode `\^12\catcode `\_12\catcode `\%12\relax}%
\providecommand \@@startlink[1]{}%
\providecommand \@@endlink[0]{}%
\providecommand \url  [0]{\begingroup\@sanitize@url \@url }%
\providecommand \@url [1]{\endgroup\@href {#1}{\urlprefix }}%
\providecommand \urlprefix  [0]{URL }%
\providecommand \Eprint [0]{\href }%
\providecommand \doibase [0]{http://dx.doi.org/}%
\providecommand \selectlanguage [0]{\@gobble}%
\providecommand \bibinfo  [0]{\@secondoftwo}%
\providecommand \bibfield  [0]{\@secondoftwo}%
\providecommand \translation [1]{[#1]}%
\providecommand \BibitemOpen [0]{}%
\providecommand \bibitemStop [0]{}%
\providecommand \bibitemNoStop [0]{.\EOS\space}%
\providecommand \EOS [0]{\spacefactor3000\relax}%
\providecommand \BibitemShut  [1]{\csname bibitem#1\endcsname}%
\let\auto@bib@innerbib\@empty
\bibitem [{\citenamefont {Cirac}\ and\ \citenamefont
  {Zoller}(1995)}]{PhysRevLett.74.4091}%
  \BibitemOpen
  \bibfield  {author} {\bibinfo {author} {\bibfnamefont {J.~I.}\ \bibnamefont
  {Cirac}}\ and\ \bibinfo {author} {\bibfnamefont {P.}~\bibnamefont {Zoller}},\
  }\href {\doibase 10.1103/PhysRevLett.74.4091} {\bibfield  {journal} {\bibinfo
   {journal} {Phys. Rev. Lett.}\ }\textbf {\bibinfo {volume} {74}},\ \bibinfo
  {pages} {4091} (\bibinfo {year} {1995})}\BibitemShut {NoStop}%
\bibitem [{\citenamefont {Blatt}\ and\ \citenamefont
  {Roos}(2012)}]{Blatt:2012aa}%
  \BibitemOpen
  \bibfield  {author} {\bibinfo {author} {\bibfnamefont {R.}~\bibnamefont
  {Blatt}}\ and\ \bibinfo {author} {\bibfnamefont {C.~F.}\ \bibnamefont
  {Roos}},\ }\href {http://dx.doi.org/10.1038/nphys2252} {\bibfield  {journal}
  {\bibinfo  {journal} {Nature Physics}\ }\textbf {\bibinfo {volume} {8}},\
  \bibinfo {pages} {277 EP } (\bibinfo {year} {2012})}\BibitemShut {NoStop}%
\bibitem [{\citenamefont {Wineland}\ \emph {et~al.}(1992)\citenamefont
  {Wineland}, \citenamefont {Bollinger}, \citenamefont {Itano}, \citenamefont
  {Moore},\ and\ \citenamefont {Heinzen}}]{PhysRevA.46.R6797}%
  \BibitemOpen
  \bibfield  {author} {\bibinfo {author} {\bibfnamefont {D.~J.}\ \bibnamefont
  {Wineland}}, \bibinfo {author} {\bibfnamefont {J.~J.}\ \bibnamefont
  {Bollinger}}, \bibinfo {author} {\bibfnamefont {W.~M.}\ \bibnamefont
  {Itano}}, \bibinfo {author} {\bibfnamefont {F.~L.}\ \bibnamefont {Moore}}, \
  and\ \bibinfo {author} {\bibfnamefont {D.~J.}\ \bibnamefont {Heinzen}},\
  }\href {\doibase 10.1103/PhysRevA.46.R6797} {\bibfield  {journal} {\bibinfo
  {journal} {Phys. Rev. A}\ }\textbf {\bibinfo {volume} {46}},\ \bibinfo
  {pages} {R6797} (\bibinfo {year} {1992})}\BibitemShut {NoStop}%
\bibitem [{\citenamefont {Ballance}\ \emph {et~al.}(2016)\citenamefont
  {Ballance}, \citenamefont {Harty}, \citenamefont {Linke}, \citenamefont
  {Sepiol},\ and\ \citenamefont {Lucas}}]{Ballance:2016}%
  \BibitemOpen
  \bibfield  {author} {\bibinfo {author} {\bibfnamefont {C.~J.}\ \bibnamefont
  {Ballance}}, \bibinfo {author} {\bibfnamefont {T.~P.}\ \bibnamefont {Harty}},
  \bibinfo {author} {\bibfnamefont {N.~M.}\ \bibnamefont {Linke}}, \bibinfo
  {author} {\bibfnamefont {M.~A.}\ \bibnamefont {Sepiol}}, \ and\ \bibinfo
  {author} {\bibfnamefont {D.~M.}\ \bibnamefont {Lucas}},\ }\href {\doibase
  10.1103/PhysRevLett.117.060504} {\bibfield  {journal} {\bibinfo  {journal}
  {Phys. Rev. Lett.}\ }\textbf {\bibinfo {volume} {117}},\ \bibinfo {pages}
  {060504} (\bibinfo {year} {2016})}\BibitemShut {NoStop}%
\bibitem [{\citenamefont {Gaebler}\ \emph {et~al.}(2016)\citenamefont
  {Gaebler}, \citenamefont {Tan}, \citenamefont {Lin}, \citenamefont {Wan},
  \citenamefont {Bowler}, \citenamefont {Keith}, \citenamefont {Glancy},
  \citenamefont {Coakley}, \citenamefont {Knill}, \citenamefont {Leibfried},\
  and\ \citenamefont {Wineland}}]{Gaebler:2016aa}%
  \BibitemOpen
  \bibfield  {author} {\bibinfo {author} {\bibfnamefont {J.~P.}\ \bibnamefont
  {Gaebler}}, \bibinfo {author} {\bibfnamefont {T.~R.}\ \bibnamefont {Tan}},
  \bibinfo {author} {\bibfnamefont {Y.}~\bibnamefont {Lin}}, \bibinfo {author}
  {\bibfnamefont {Y.}~\bibnamefont {Wan}}, \bibinfo {author} {\bibfnamefont
  {R.}~\bibnamefont {Bowler}}, \bibinfo {author} {\bibfnamefont {A.~C.}\
  \bibnamefont {Keith}}, \bibinfo {author} {\bibfnamefont {S.}~\bibnamefont
  {Glancy}}, \bibinfo {author} {\bibfnamefont {K.}~\bibnamefont {Coakley}},
  \bibinfo {author} {\bibfnamefont {E.}~\bibnamefont {Knill}}, \bibinfo
  {author} {\bibfnamefont {D.}~\bibnamefont {Leibfried}}, \ and\ \bibinfo
  {author} {\bibfnamefont {D.~J.}\ \bibnamefont {Wineland}},\ }\href {\doibase
  10.1103/PhysRevLett.117.060505} {\bibfield  {journal} {\bibinfo  {journal}
  {Phys. Rev. Lett.}\ }\textbf {\bibinfo {volume} {117}},\ \bibinfo {pages}
  {060505} (\bibinfo {year} {2016})}\BibitemShut {NoStop}%
\bibitem [{\citenamefont {Zhang}\ \emph {et~al.}(2017)\citenamefont {Zhang},
  \citenamefont {Pagano}, \citenamefont {Hess}, \citenamefont {Kyprianidis},
  \citenamefont {Becker}, \citenamefont {Kaplan}, \citenamefont {Gorshkov},
  \citenamefont {Gong},\ and\ \citenamefont {Monroe}}]{Zhang_2017}%
  \BibitemOpen
  \bibfield  {author} {\bibinfo {author} {\bibfnamefont {J.}~\bibnamefont
  {Zhang}}, \bibinfo {author} {\bibfnamefont {G.}~\bibnamefont {Pagano}},
  \bibinfo {author} {\bibfnamefont {P.~W.}\ \bibnamefont {Hess}}, \bibinfo
  {author} {\bibfnamefont {A.}~\bibnamefont {Kyprianidis}}, \bibinfo {author}
  {\bibfnamefont {P.}~\bibnamefont {Becker}}, \bibinfo {author} {\bibfnamefont
  {H.}~\bibnamefont {Kaplan}}, \bibinfo {author} {\bibfnamefont {A.~V.}\
  \bibnamefont {Gorshkov}}, \bibinfo {author} {\bibfnamefont {Z.~X.}\
  \bibnamefont {Gong}}, \ and\ \bibinfo {author} {\bibfnamefont
  {C.}~\bibnamefont {Monroe}},\ }\href {http://dx.doi.org/10.1038/nature24654}
  {\bibfield  {journal} {\bibinfo  {journal} {Nature}\ }\textbf {\bibinfo
  {volume} {551}},\ \bibinfo {pages} {601 EP } (\bibinfo {year}
  {2017})}\BibitemShut {NoStop}%
\bibitem [{\citenamefont {Bohnet}\ \emph {et~al.}(2016)\citenamefont {Bohnet},
  \citenamefont {Sawyer}, \citenamefont {Britton}, \citenamefont {Wall},
  \citenamefont {Rey}, \citenamefont {Foss-feig},\ and\ \citenamefont
  {Bollinger}}]{Bohnet2016}%
  \BibitemOpen
  \bibfield  {author} {\bibinfo {author} {\bibfnamefont {J.~G.}\ \bibnamefont
  {Bohnet}}, \bibinfo {author} {\bibfnamefont {B.~C.}\ \bibnamefont {Sawyer}},
  \bibinfo {author} {\bibfnamefont {J.~W.}\ \bibnamefont {Britton}}, \bibinfo
  {author} {\bibfnamefont {M.~L.}\ \bibnamefont {Wall}}, \bibinfo {author}
  {\bibfnamefont {A.~M.}\ \bibnamefont {Rey}}, \bibinfo {author} {\bibfnamefont
  {M.}~\bibnamefont {Foss-feig}}, \ and\ \bibinfo {author} {\bibfnamefont
  {J.~J.}\ \bibnamefont {Bollinger}},\ }\href@noop {} {\bibfield  {journal}
  {\bibinfo  {journal} {Science}\ }\textbf {\bibinfo {volume} {352}},\ \bibinfo
  {pages} {1297} (\bibinfo {year} {2016})}\BibitemShut {NoStop}%
\bibitem [{\citenamefont {Monroe}\ \emph {et~al.}(1995)\citenamefont {Monroe},
  \citenamefont {Meekhof}, \citenamefont {King}, \citenamefont {Itano},\ and\
  \citenamefont {Wineland}}]{PhysRevLett.75.4714}%
  \BibitemOpen
  \bibfield  {author} {\bibinfo {author} {\bibfnamefont {C.}~\bibnamefont
  {Monroe}}, \bibinfo {author} {\bibfnamefont {D.~M.}\ \bibnamefont {Meekhof}},
  \bibinfo {author} {\bibfnamefont {B.~E.}\ \bibnamefont {King}}, \bibinfo
  {author} {\bibfnamefont {W.~M.}\ \bibnamefont {Itano}}, \ and\ \bibinfo
  {author} {\bibfnamefont {D.~J.}\ \bibnamefont {Wineland}},\ }\href {\doibase
  10.1103/PhysRevLett.75.4714} {\bibfield  {journal} {\bibinfo  {journal}
  {Phys. Rev. Lett.}\ }\textbf {\bibinfo {volume} {75}},\ \bibinfo {pages}
  {4714} (\bibinfo {year} {1995})}\BibitemShut {NoStop}%
\bibitem [{\citenamefont {Ospelkaus}\ \emph {et~al.}(2011)\citenamefont
  {Ospelkaus}, \citenamefont {Warring}, \citenamefont {Colombe}, \citenamefont
  {Brown}, \citenamefont {Amini}, \citenamefont {Leibfried},\ and\
  \citenamefont {Wineland}}]{maggrad1}%
  \BibitemOpen
  \bibfield  {author} {\bibinfo {author} {\bibfnamefont {C.}~\bibnamefont
  {Ospelkaus}}, \bibinfo {author} {\bibfnamefont {U.}~\bibnamefont {Warring}},
  \bibinfo {author} {\bibfnamefont {Y.}~\bibnamefont {Colombe}}, \bibinfo
  {author} {\bibfnamefont {K.}~\bibnamefont {Brown}}, \bibinfo {author}
  {\bibfnamefont {J.}~\bibnamefont {Amini}}, \bibinfo {author} {\bibfnamefont
  {D.}~\bibnamefont {Leibfried}}, \ and\ \bibinfo {author} {\bibfnamefont
  {D.}~\bibnamefont {Wineland}},\ }\href@noop {} {\bibfield  {journal}
  {\bibinfo  {journal} {Nature}\ }\textbf {\bibinfo {volume} {476}},\ \bibinfo
  {pages} {181} (\bibinfo {year} {2011})}\BibitemShut {NoStop}%
\bibitem [{\citenamefont {Harty}\ \emph {et~al.}(2016)\citenamefont {Harty},
  \citenamefont {Sepiol}, \citenamefont {Allcock}, \citenamefont {Ballance},
  \citenamefont {Tarlton},\ and\ \citenamefont {Lucas}}]{maggrad2}%
  \BibitemOpen
  \bibfield  {author} {\bibinfo {author} {\bibfnamefont {T.~P.}\ \bibnamefont
  {Harty}}, \bibinfo {author} {\bibfnamefont {M.~A.}\ \bibnamefont {Sepiol}},
  \bibinfo {author} {\bibfnamefont {D.~T.~C.}\ \bibnamefont {Allcock}},
  \bibinfo {author} {\bibfnamefont {C.~J.}\ \bibnamefont {Ballance}}, \bibinfo
  {author} {\bibfnamefont {J.~E.}\ \bibnamefont {Tarlton}}, \ and\ \bibinfo
  {author} {\bibfnamefont {D.~M.}\ \bibnamefont {Lucas}},\ }\href {\doibase
  10.1103/PhysRevLett.117.140501} {\bibfield  {journal} {\bibinfo  {journal}
  {Phys. Rev. Lett.}\ }\textbf {\bibinfo {volume} {117}},\ \bibinfo {pages}
  {140501} (\bibinfo {year} {2016})}\BibitemShut {NoStop}%
\bibitem [{\citenamefont {Ozeri}\ \emph {et~al.}(2007)\citenamefont {Ozeri},
  \citenamefont {Itano}, \citenamefont {Blakestad}, \citenamefont {Britton},
  \citenamefont {Chiaverini}, \citenamefont {Jost}, \citenamefont {Langer},
  \citenamefont {Leibfried}, \citenamefont {Reichle}, \citenamefont {Seidelin},
  \citenamefont {Wesenberg},\ and\ \citenamefont
  {Wineland}}]{PhysRevA.75.042329}%
  \BibitemOpen
  \bibfield  {author} {\bibinfo {author} {\bibfnamefont {R.}~\bibnamefont
  {Ozeri}}, \bibinfo {author} {\bibfnamefont {W.~M.}\ \bibnamefont {Itano}},
  \bibinfo {author} {\bibfnamefont {R.~B.}\ \bibnamefont {Blakestad}}, \bibinfo
  {author} {\bibfnamefont {J.}~\bibnamefont {Britton}}, \bibinfo {author}
  {\bibfnamefont {J.}~\bibnamefont {Chiaverini}}, \bibinfo {author}
  {\bibfnamefont {J.~D.}\ \bibnamefont {Jost}}, \bibinfo {author}
  {\bibfnamefont {C.}~\bibnamefont {Langer}}, \bibinfo {author} {\bibfnamefont
  {D.}~\bibnamefont {Leibfried}}, \bibinfo {author} {\bibfnamefont
  {R.}~\bibnamefont {Reichle}}, \bibinfo {author} {\bibfnamefont
  {S.}~\bibnamefont {Seidelin}}, \bibinfo {author} {\bibfnamefont {J.~H.}\
  \bibnamefont {Wesenberg}}, \ and\ \bibinfo {author} {\bibfnamefont {D.~J.}\
  \bibnamefont {Wineland}},\ }\href {\doibase 10.1103/PhysRevA.75.042329}
  {\bibfield  {journal} {\bibinfo  {journal} {Phys. Rev. A}\ }\textbf {\bibinfo
  {volume} {75}},\ \bibinfo {pages} {042329} (\bibinfo {year}
  {2007})}\BibitemShut {NoStop}%
\bibitem [{\citenamefont {Uys}\ \emph {et~al.}(2010)\citenamefont {Uys},
  \citenamefont {Biercuk}, \citenamefont {VanDevender}, \citenamefont
  {Ospelkaus}, \citenamefont {Meiser}, \citenamefont {Ozeri},\ and\
  \citenamefont {Bollinger}}]{PhysRevLett.105.200401}%
  \BibitemOpen
  \bibfield  {author} {\bibinfo {author} {\bibfnamefont {H.}~\bibnamefont
  {Uys}}, \bibinfo {author} {\bibfnamefont {M.~J.}\ \bibnamefont {Biercuk}},
  \bibinfo {author} {\bibfnamefont {A.~P.}\ \bibnamefont {VanDevender}},
  \bibinfo {author} {\bibfnamefont {C.}~\bibnamefont {Ospelkaus}}, \bibinfo
  {author} {\bibfnamefont {D.}~\bibnamefont {Meiser}}, \bibinfo {author}
  {\bibfnamefont {R.}~\bibnamefont {Ozeri}}, \ and\ \bibinfo {author}
  {\bibfnamefont {J.~J.}\ \bibnamefont {Bollinger}},\ }\href {\doibase
  10.1103/PhysRevLett.105.200401} {\bibfield  {journal} {\bibinfo  {journal}
  {Phys. Rev. Lett.}\ }\textbf {\bibinfo {volume} {105}},\ \bibinfo {pages}
  {200401} (\bibinfo {year} {2010})}\BibitemShut {NoStop}%
\bibitem [{\citenamefont {Brown}\ and\ \citenamefont
  {Brown}(2018)}]{PhysRevA.97.052301}%
  \BibitemOpen
  \bibfield  {author} {\bibinfo {author} {\bibfnamefont {N.~C.}\ \bibnamefont
  {Brown}}\ and\ \bibinfo {author} {\bibfnamefont {K.~R.}\ \bibnamefont
  {Brown}},\ }\href {\doibase 10.1103/PhysRevA.97.052301} {\bibfield  {journal}
  {\bibinfo  {journal} {Phys. Rev. A}\ }\textbf {\bibinfo {volume} {97}},\
  \bibinfo {pages} {052301} (\bibinfo {year} {2018})}\BibitemShut {NoStop}%
\bibitem [{\citenamefont {Choi}\ \emph {et~al.}(2014)\citenamefont {Choi},
  \citenamefont {Debnath}, \citenamefont {Manning}, \citenamefont {Figgatt},
  \citenamefont {Gong}, \citenamefont {Duan},\ and\ \citenamefont
  {Monroe}}]{PhysRevLett.112.190502}%
  \BibitemOpen
  \bibfield  {author} {\bibinfo {author} {\bibfnamefont {T.}~\bibnamefont
  {Choi}}, \bibinfo {author} {\bibfnamefont {S.}~\bibnamefont {Debnath}},
  \bibinfo {author} {\bibfnamefont {T.~A.}\ \bibnamefont {Manning}}, \bibinfo
  {author} {\bibfnamefont {C.}~\bibnamefont {Figgatt}}, \bibinfo {author}
  {\bibfnamefont {Z.-X.}\ \bibnamefont {Gong}}, \bibinfo {author}
  {\bibfnamefont {L.-M.}\ \bibnamefont {Duan}}, \ and\ \bibinfo {author}
  {\bibfnamefont {C.}~\bibnamefont {Monroe}},\ }\href {\doibase
  10.1103/PhysRevLett.112.190502} {\bibfield  {journal} {\bibinfo  {journal}
  {Phys. Rev. Lett.}\ }\textbf {\bibinfo {volume} {112}},\ \bibinfo {pages}
  {190502} (\bibinfo {year} {2014})}\BibitemShut {NoStop}%
\bibitem [{\citenamefont {Green}\ and\ \citenamefont
  {Biercuk}(2015)}]{PhysRevLett.114.120502}%
  \BibitemOpen
  \bibfield  {author} {\bibinfo {author} {\bibfnamefont {T.~J.}\ \bibnamefont
  {Green}}\ and\ \bibinfo {author} {\bibfnamefont {M.~J.}\ \bibnamefont
  {Biercuk}},\ }\href {\doibase 10.1103/PhysRevLett.114.120502} {\bibfield
  {journal} {\bibinfo  {journal} {Phys. Rev. Lett.}\ }\textbf {\bibinfo
  {volume} {114}},\ \bibinfo {pages} {120502} (\bibinfo {year}
  {2015})}\BibitemShut {NoStop}%
\bibitem [{\citenamefont {Leung}\ \emph {et~al.}(2018)\citenamefont {Leung},
  \citenamefont {Landsman}, \citenamefont {Figgatt}, \citenamefont {Linke},
  \citenamefont {Monroe},\ and\ \citenamefont
  {Brown}}]{PhysRevLett.120.020501}%
  \BibitemOpen
  \bibfield  {author} {\bibinfo {author} {\bibfnamefont {P.~H.}\ \bibnamefont
  {Leung}}, \bibinfo {author} {\bibfnamefont {K.~A.}\ \bibnamefont {Landsman}},
  \bibinfo {author} {\bibfnamefont {C.}~\bibnamefont {Figgatt}}, \bibinfo
  {author} {\bibfnamefont {N.~M.}\ \bibnamefont {Linke}}, \bibinfo {author}
  {\bibfnamefont {C.}~\bibnamefont {Monroe}}, \ and\ \bibinfo {author}
  {\bibfnamefont {K.~R.}\ \bibnamefont {Brown}},\ }\href {\doibase
  10.1103/PhysRevLett.120.020501} {\bibfield  {journal} {\bibinfo  {journal}
  {Phys. Rev. Lett.}\ }\textbf {\bibinfo {volume} {120}},\ \bibinfo {pages}
  {020501} (\bibinfo {year} {2018})}\BibitemShut {NoStop}%
\bibitem [{\citenamefont {Heinzen}\ and\ \citenamefont
  {Wineland}(1990)}]{Heinzen1990}%
  \BibitemOpen
  \bibfield  {author} {\bibinfo {author} {\bibfnamefont {D.~J.}\ \bibnamefont
  {Heinzen}}\ and\ \bibinfo {author} {\bibfnamefont {D.~J.}\ \bibnamefont
  {Wineland}},\ }\href {\doibase 10.1103/PhysRevA.42.2977} {\bibfield
  {journal} {\bibinfo  {journal} {Physical Review A}\ }\textbf {\bibinfo
  {volume} {42}},\ \bibinfo {pages} {2977} (\bibinfo {year}
  {1990})}\BibitemShut {NoStop}%
\bibitem [{\citenamefont {Babadi}\ \emph {et~al.}(2017)\citenamefont {Babadi},
  \citenamefont {Knap}, \citenamefont {Martin}, \citenamefont {Refael},\ and\
  \citenamefont {Demler}}]{Babadi2017}%
  \BibitemOpen
  \bibfield  {author} {\bibinfo {author} {\bibfnamefont {M.}~\bibnamefont
  {Babadi}}, \bibinfo {author} {\bibfnamefont {M.}~\bibnamefont {Knap}},
  \bibinfo {author} {\bibfnamefont {I.}~\bibnamefont {Martin}}, \bibinfo
  {author} {\bibfnamefont {G.}~\bibnamefont {Refael}}, \ and\ \bibinfo {author}
  {\bibfnamefont {E.}~\bibnamefont {Demler}},\ }\href {\doibase
  10.1103/PhysRevB.96.014512} {\bibfield  {journal} {\bibinfo  {journal} {Phys.
  Rev. B}\ }\textbf {\bibinfo {volume} {96}},\ \bibinfo {pages} {014512}
  (\bibinfo {year} {2017})}\BibitemShut {NoStop}%
\bibitem [{\citenamefont {Lemonde}\ \emph {et~al.}(2016)\citenamefont
  {Lemonde}, \citenamefont {Didier},\ and\ \citenamefont
  {Clerk}}]{lemonde2016enhanced}%
  \BibitemOpen
  \bibfield  {author} {\bibinfo {author} {\bibfnamefont {M.-A.}\ \bibnamefont
  {Lemonde}}, \bibinfo {author} {\bibfnamefont {N.}~\bibnamefont {Didier}}, \
  and\ \bibinfo {author} {\bibfnamefont {A.~A.}\ \bibnamefont {Clerk}},\
  }\href@noop {} {\bibfield  {journal} {\bibinfo  {journal} {Nature
  communications}\ }\textbf {\bibinfo {volume} {7}},\ \bibinfo {pages} {11338}
  (\bibinfo {year} {2016})}\BibitemShut {NoStop}%
\bibitem [{\citenamefont {Zeytinoglu}\ \emph {et~al.}(2017)\citenamefont
  {Zeytinoglu}, \citenamefont {Imamoglu},\ and\ \citenamefont
  {Huber}}]{PhysRevX.7.021041}%
  \BibitemOpen
  \bibfield  {author} {\bibinfo {author} {\bibfnamefont {S.}~\bibnamefont
  {Zeytinoglu}}, \bibinfo {author} {\bibfnamefont {A.}~\bibnamefont
  {Imamoglu}}, \ and\ \bibinfo {author} {\bibfnamefont {S.}~\bibnamefont
  {Huber}},\ }\href {\doibase 10.1103/PhysRevX.7.021041} {\bibfield  {journal}
  {\bibinfo  {journal} {Phys. Rev. X}\ }\textbf {\bibinfo {volume} {7}},\
  \bibinfo {pages} {021041} (\bibinfo {year} {2017})}\BibitemShut {NoStop}%
\bibitem [{\citenamefont {Arenz}\ \emph {et~al.}(2018)\citenamefont {Arenz},
  \citenamefont {Bondar}, \citenamefont {Burgarth}, \citenamefont {Cormick},\
  and\ \citenamefont {Rabitz}}]{arenz2018hamiltonian}%
  \BibitemOpen
  \bibfield  {author} {\bibinfo {author} {\bibfnamefont {C.}~\bibnamefont
  {Arenz}}, \bibinfo {author} {\bibfnamefont {D.~I.}\ \bibnamefont {Bondar}},
  \bibinfo {author} {\bibfnamefont {D.}~\bibnamefont {Burgarth}}, \bibinfo
  {author} {\bibfnamefont {C.}~\bibnamefont {Cormick}}, \ and\ \bibinfo
  {author} {\bibfnamefont {H.}~\bibnamefont {Rabitz}},\ }\href@noop {}
  {\bibfield  {journal} {\bibinfo  {journal} {arXiv preprint arXiv:1806.00444}\
  } (\bibinfo {year} {2018})}\BibitemShut {NoStop}%
\bibitem [{\citenamefont {S\o{}rensen}\ and\ \citenamefont
  {M\o{}lmer}(2000)}]{Sorensen00}%
  \BibitemOpen
  \bibfield  {author} {\bibinfo {author} {\bibfnamefont {A.}~\bibnamefont
  {S\o{}rensen}}\ and\ \bibinfo {author} {\bibfnamefont {K.}~\bibnamefont
  {M\o{}lmer}},\ }\href {\doibase 10.1103/PhysRevA.62.022311} {\bibfield
  {journal} {\bibinfo  {journal} {Phys. Rev. A}\ }\textbf {\bibinfo {volume}
  {62}},\ \bibinfo {pages} {022311} (\bibinfo {year} {2000})}\BibitemShut
  {NoStop}%
\bibitem [{\citenamefont {Leibfried}\ \emph {et~al.}(2003)\citenamefont
  {Leibfried}, \citenamefont {DeMarco}, \citenamefont {Meyer}, \citenamefont
  {Lucas}, \citenamefont {Barrett}, \citenamefont {Britton}, \citenamefont
  {Itano}, \citenamefont {Jelenkovi{\'c}}, \citenamefont {Langer},
  \citenamefont {Rosenband},\ and\ \citenamefont {Wineland}}]{Leibfried03}%
  \BibitemOpen
  \bibfield  {author} {\bibinfo {author} {\bibfnamefont {D.}~\bibnamefont
  {Leibfried}}, \bibinfo {author} {\bibfnamefont {B.}~\bibnamefont {DeMarco}},
  \bibinfo {author} {\bibfnamefont {V.}~\bibnamefont {Meyer}}, \bibinfo
  {author} {\bibfnamefont {D.}~\bibnamefont {Lucas}}, \bibinfo {author}
  {\bibfnamefont {M.}~\bibnamefont {Barrett}}, \bibinfo {author} {\bibfnamefont
  {J.}~\bibnamefont {Britton}}, \bibinfo {author} {\bibfnamefont {W.~M.}\
  \bibnamefont {Itano}}, \bibinfo {author} {\bibfnamefont {B.}~\bibnamefont
  {Jelenkovi{\'c}}}, \bibinfo {author} {\bibfnamefont {C.}~\bibnamefont
  {Langer}}, \bibinfo {author} {\bibfnamefont {T.}~\bibnamefont {Rosenband}}, \
  and\ \bibinfo {author} {\bibfnamefont {D.~J.}\ \bibnamefont {Wineland}},\
  }\href {http://dx.doi.org/10.1038/nature01492} {\bibfield  {journal}
  {\bibinfo  {journal} {Nature}\ }\textbf {\bibinfo {volume} {422}},\ \bibinfo
  {pages} {412} (\bibinfo {year} {2003})}\BibitemShut {NoStop}%
\bibitem [{\citenamefont {Garc\'{\i}a-Ripoll}\ \emph
  {et~al.}(2005)\citenamefont {Garc\'{\i}a-Ripoll}, \citenamefont {Zoller},\
  and\ \citenamefont {Cirac}}]{G-Ripoll05}%
  \BibitemOpen
  \bibfield  {author} {\bibinfo {author} {\bibfnamefont {J.~J.}\ \bibnamefont
  {Garc\'{\i}a-Ripoll}}, \bibinfo {author} {\bibfnamefont {P.}~\bibnamefont
  {Zoller}}, \ and\ \bibinfo {author} {\bibfnamefont {J.~I.}\ \bibnamefont
  {Cirac}},\ }\href {\doibase 10.1103/PhysRevA.71.062309} {\bibfield  {journal}
  {\bibinfo  {journal} {Phys. Rev. A}\ }\textbf {\bibinfo {volume} {71}},\
  \bibinfo {pages} {062309} (\bibinfo {year} {2005})}\BibitemShut {NoStop}%
\bibitem [{SM()}]{SM}%
  \BibitemOpen
  \href@noop {} {}\bibinfo {organization} {See Supplementary Materials for
  technical details, which includes Refs. \cite{milburn_phase_gate,
  DylewskyPRA}}\BibitemShut {NoStop}%
\bibitem [{\citenamefont {Ospelkaus}\ \emph {et~al.}(2008)\citenamefont
  {Ospelkaus}, \citenamefont {Langer}, \citenamefont {Amini}, \citenamefont
  {Brown}, \citenamefont {Leibfried},\ and\ \citenamefont
  {Wineland}}]{PhysRevLett.101.090502}%
  \BibitemOpen
  \bibfield  {author} {\bibinfo {author} {\bibfnamefont {C.}~\bibnamefont
  {Ospelkaus}}, \bibinfo {author} {\bibfnamefont {C.~E.}\ \bibnamefont
  {Langer}}, \bibinfo {author} {\bibfnamefont {J.~M.}\ \bibnamefont {Amini}},
  \bibinfo {author} {\bibfnamefont {K.~R.}\ \bibnamefont {Brown}}, \bibinfo
  {author} {\bibfnamefont {D.}~\bibnamefont {Leibfried}}, \ and\ \bibinfo
  {author} {\bibfnamefont {D.~J.}\ \bibnamefont {Wineland}},\ }\href {\doibase
  10.1103/PhysRevLett.101.090502} {\bibfield  {journal} {\bibinfo  {journal}
  {Phys. Rev. Lett.}\ }\textbf {\bibinfo {volume} {101}},\ \bibinfo {pages}
  {090502} (\bibinfo {year} {2008})}\BibitemShut {NoStop}%
\bibitem [{\citenamefont {Britton}\ \emph {et~al.}(2012)\citenamefont
  {Britton}, \citenamefont {Sawyer}, \citenamefont {Keith}, \citenamefont
  {Wang}, \citenamefont {Freericks}, \citenamefont {Uys}, \citenamefont
  {Biercuk},\ and\ \citenamefont {Bollinger}}]{Britton2012}%
  \BibitemOpen
  \bibfield  {author} {\bibinfo {author} {\bibfnamefont {J.~W.}\ \bibnamefont
  {Britton}}, \bibinfo {author} {\bibfnamefont {B.~C.}\ \bibnamefont {Sawyer}},
  \bibinfo {author} {\bibfnamefont {A.~C.}\ \bibnamefont {Keith}}, \bibinfo
  {author} {\bibfnamefont {C.-C.~J.}\ \bibnamefont {Wang}}, \bibinfo {author}
  {\bibfnamefont {J.~K.}\ \bibnamefont {Freericks}}, \bibinfo {author}
  {\bibfnamefont {H.}~\bibnamefont {Uys}}, \bibinfo {author} {\bibfnamefont
  {M.~J.}\ \bibnamefont {Biercuk}}, \ and\ \bibinfo {author} {\bibfnamefont
  {J.~J.}\ \bibnamefont {Bollinger}},\ }\href {\doibase 10.1038/nature10981}
  {\bibfield  {journal} {\bibinfo  {journal} {Nature}\ }\textbf {\bibinfo
  {volume} {484}},\ \bibinfo {pages} {489} (\bibinfo {year} {2012})},\ \Eprint
  {http://arxiv.org/abs/1204.5789} {arXiv:1204.5789} \BibitemShut {NoStop}%
\bibitem [{\citenamefont {James}(1998)}]{james1998quantum}%
  \BibitemOpen
  \bibfield  {author} {\bibinfo {author} {\bibfnamefont {D.~F.}\ \bibnamefont
  {James}},\ }\href@noop {} {\bibfield  {journal} {\bibinfo  {journal} {Applied
  Physics B: Lasers and Optics}\ }\textbf {\bibinfo {volume} {66}},\ \bibinfo
  {pages} {181} (\bibinfo {year} {1998})}\BibitemShut {NoStop}%
\bibitem [{\citenamefont {Porras}\ and\ \citenamefont
  {Cirac}(2004)}]{PhysRevLett.92.207901}%
  \BibitemOpen
  \bibfield  {author} {\bibinfo {author} {\bibfnamefont {D.}~\bibnamefont
  {Porras}}\ and\ \bibinfo {author} {\bibfnamefont {J.~I.}\ \bibnamefont
  {Cirac}},\ }\href {\doibase 10.1103/PhysRevLett.92.207901} {\bibfield
  {journal} {\bibinfo  {journal} {Phys. Rev. Lett.}\ }\textbf {\bibinfo
  {volume} {92}},\ \bibinfo {pages} {207901} (\bibinfo {year}
  {2004})}\BibitemShut {NoStop}%
\bibitem [{\citenamefont {Kim}\ \emph {et~al.}(2009)\citenamefont {Kim},
  \citenamefont {Chang}, \citenamefont {Islam}, \citenamefont {Korenblit},
  \citenamefont {Duan},\ and\ \citenamefont {Monroe}}]{Kim2009}%
  \BibitemOpen
  \bibfield  {author} {\bibinfo {author} {\bibfnamefont {K.}~\bibnamefont
  {Kim}}, \bibinfo {author} {\bibfnamefont {M.}~\bibnamefont {Chang}}, \bibinfo
  {author} {\bibfnamefont {R.}~\bibnamefont {Islam}}, \bibinfo {author}
  {\bibfnamefont {S.}~\bibnamefont {Korenblit}}, \bibinfo {author}
  {\bibfnamefont {L.}~\bibnamefont {Duan}}, \ and\ \bibinfo {author}
  {\bibfnamefont {C.}~\bibnamefont {Monroe}},\ }\href {\doibase
  10.1103/PhysRevLett.103.120502} {\bibfield  {journal} {\bibinfo  {journal}
  {Physical Review Letters}\ }\textbf {\bibinfo {volume} {103}},\ \bibinfo
  {pages} {120502} (\bibinfo {year} {2009})}\BibitemShut {NoStop}%
\bibitem [{Note1()}]{Note1}%
  \BibitemOpen
  \bibinfo {note} {In general, the condition is $|g|<|\delta _m|$. In this
  work, we assume $0<g<\delta _m$ for simplicity.}\BibitemShut {Stop}%
\bibitem [{\citenamefont {Carmichael}\ \emph {et~al.}(1984)\citenamefont
  {Carmichael}, \citenamefont {Milburn},\ and\ \citenamefont
  {Walls}}]{carmichael1984squeezing}%
  \BibitemOpen
  \bibfield  {author} {\bibinfo {author} {\bibfnamefont {H.}~\bibnamefont
  {Carmichael}}, \bibinfo {author} {\bibfnamefont {G.}~\bibnamefont {Milburn}},
  \ and\ \bibinfo {author} {\bibfnamefont {D.}~\bibnamefont {Walls}},\
  }\href@noop {} {\bibfield  {journal} {\bibinfo  {journal} {Journal of Physics
  A: Mathematical and General}\ }\textbf {\bibinfo {volume} {17}},\ \bibinfo
  {pages} {469} (\bibinfo {year} {1984})}\BibitemShut {NoStop}%
\bibitem [{Note2()}]{Note2}%
  \BibitemOpen
  \bibinfo {note} {The geometric phase is obtained by numerically solving
  Eq.\protect \tmspace +\thinmuskip {.1667em}\protect \textup {\hbox
  {\mathsurround \z@ \protect \normalfont (\ignorespaces \ref
  {eq:BmodeH}\unskip \@@italiccorr )}}, and extracting the area enclosed by the
  phase-space trajectories in the interaction picture of the total quadratic
  terms.}\BibitemShut {Stop}%
\bibitem [{\citenamefont {T\'oth}\ \emph {et~al.}(2007)\citenamefont {T\'oth},
  \citenamefont {Knapp}, \citenamefont {G\"uhne},\ and\ \citenamefont
  {Briegel}}]{PhysRevLett.99.250405}%
  \BibitemOpen
  \bibfield  {author} {\bibinfo {author} {\bibfnamefont {G.}~\bibnamefont
  {T\'oth}}, \bibinfo {author} {\bibfnamefont {C.}~\bibnamefont {Knapp}},
  \bibinfo {author} {\bibfnamefont {O.}~\bibnamefont {G\"uhne}}, \ and\
  \bibinfo {author} {\bibfnamefont {H.~J.}\ \bibnamefont {Briegel}},\ }\href
  {\doibase 10.1103/PhysRevLett.99.250405} {\bibfield  {journal} {\bibinfo
  {journal} {Phys. Rev. Lett.}\ }\textbf {\bibinfo {volume} {99}},\ \bibinfo
  {pages} {250405} (\bibinfo {year} {2007})}\BibitemShut {NoStop}%
\bibitem [{\citenamefont {Ma}\ \emph {et~al.}(2011)\citenamefont {Ma},
  \citenamefont {Wang}, \citenamefont {Sun},\ and\ \citenamefont
  {Nori}}]{ma2011quantum}%
  \BibitemOpen
  \bibfield  {author} {\bibinfo {author} {\bibfnamefont {J.}~\bibnamefont
  {Ma}}, \bibinfo {author} {\bibfnamefont {X.}~\bibnamefont {Wang}}, \bibinfo
  {author} {\bibfnamefont {C.-P.}\ \bibnamefont {Sun}}, \ and\ \bibinfo
  {author} {\bibfnamefont {F.}~\bibnamefont {Nori}},\ }\href@noop {} {\bibfield
   {journal} {\bibinfo  {journal} {Physics Reports}\ }\textbf {\bibinfo
  {volume} {509}},\ \bibinfo {pages} {89} (\bibinfo {year} {2011})}\BibitemShut
  {NoStop}%
\bibitem [{\citenamefont {Wineland}\ \emph {et~al.}(1994)\citenamefont
  {Wineland}, \citenamefont {Bollinger}, \citenamefont {Itano},\ and\
  \citenamefont {Heinzen}}]{PhysRevA.50.67}%
  \BibitemOpen
  \bibfield  {author} {\bibinfo {author} {\bibfnamefont {D.~J.}\ \bibnamefont
  {Wineland}}, \bibinfo {author} {\bibfnamefont {J.~J.}\ \bibnamefont
  {Bollinger}}, \bibinfo {author} {\bibfnamefont {W.~M.}\ \bibnamefont
  {Itano}}, \ and\ \bibinfo {author} {\bibfnamefont {D.~J.}\ \bibnamefont
  {Heinzen}},\ }\href {\doibase 10.1103/PhysRevA.50.67} {\bibfield  {journal}
  {\bibinfo  {journal} {Phys. Rev. A}\ }\textbf {\bibinfo {volume} {50}},\
  \bibinfo {pages} {67} (\bibinfo {year} {1994})}\BibitemShut {NoStop}%
\bibitem [{\citenamefont {Kitagawa}\ and\ \citenamefont
  {Ueda}(1993)}]{PhysRevA.47.5138}%
  \BibitemOpen
  \bibfield  {author} {\bibinfo {author} {\bibfnamefont {M.}~\bibnamefont
  {Kitagawa}}\ and\ \bibinfo {author} {\bibfnamefont {M.}~\bibnamefont
  {Ueda}},\ }\href {\doibase 10.1103/PhysRevA.47.5138} {\bibfield  {journal}
  {\bibinfo  {journal} {Phys. Rev. A}\ }\textbf {\bibinfo {volume} {47}},\
  \bibinfo {pages} {5138} (\bibinfo {year} {1993})}\BibitemShut {NoStop}%
\bibitem [{\citenamefont {Foss-Feig}\ \emph {et~al.}(2013)\citenamefont
  {Foss-Feig}, \citenamefont {Hazzard}, \citenamefont {Bollinger},\ and\
  \citenamefont {Rey}}]{foss2013nonequilibrium}%
  \BibitemOpen
  \bibfield  {author} {\bibinfo {author} {\bibfnamefont {M.}~\bibnamefont
  {Foss-Feig}}, \bibinfo {author} {\bibfnamefont {K.~R.}\ \bibnamefont
  {Hazzard}}, \bibinfo {author} {\bibfnamefont {J.~J.}\ \bibnamefont
  {Bollinger}}, \ and\ \bibinfo {author} {\bibfnamefont {A.~M.}\ \bibnamefont
  {Rey}},\ }\href@noop {} {\bibfield  {journal} {\bibinfo  {journal} {Physical
  Review A}\ }\textbf {\bibinfo {volume} {87}},\ \bibinfo {pages} {042101}
  (\bibinfo {year} {2013})}\BibitemShut {NoStop}%
\bibitem [{\citenamefont {Lewis-Swan}\ \emph {et~al.}(2018)\citenamefont
  {Lewis-Swan}, \citenamefont {Norcia}, \citenamefont {Cline}, \citenamefont
  {Thompson},\ and\ \citenamefont {Rey}}]{PhysRevLett.121.070403}%
  \BibitemOpen
  \bibfield  {author} {\bibinfo {author} {\bibfnamefont {R.~J.}\ \bibnamefont
  {Lewis-Swan}}, \bibinfo {author} {\bibfnamefont {M.~A.}\ \bibnamefont
  {Norcia}}, \bibinfo {author} {\bibfnamefont {J.~R.~K.}\ \bibnamefont
  {Cline}}, \bibinfo {author} {\bibfnamefont {J.~K.}\ \bibnamefont {Thompson}},
  \ and\ \bibinfo {author} {\bibfnamefont {A.~M.}\ \bibnamefont {Rey}},\ }\href
  {\doibase 10.1103/PhysRevLett.121.070403} {\bibfield  {journal} {\bibinfo
  {journal} {Phys. Rev. Lett.}\ }\textbf {\bibinfo {volume} {121}},\ \bibinfo
  {pages} {070403} (\bibinfo {year} {2018})}\BibitemShut {NoStop}%
\bibitem [{Note3()}]{Note3}%
  \BibitemOpen
  \bibinfo {note} {For consistency with the experiment reported in Ref.\ \cite
  {Bohnet2016}, in the numerical simulations we include the effects of both
  spontaneous spin flips and additional elastic dephasing, reporting the
  combination of these two rates as $\Gamma $ in Fig.\protect \tmspace
  +\thinmuskip {.1667em}\ref {fig:ff}.}\BibitemShut {Stop}%
\bibitem [{\citenamefont {Schmiegelow}\ \emph {et~al.}(2016)\citenamefont
  {Schmiegelow}, \citenamefont {Kaufmann}, \citenamefont {Ruster},
  \citenamefont {Schulz}, \citenamefont {Kaushal}, \citenamefont {Hettrich},
  \citenamefont {Schmidt-Kaler},\ and\ \citenamefont
  {Poschinger}}]{PhysRevLett.116.033002}%
  \BibitemOpen
  \bibfield  {author} {\bibinfo {author} {\bibfnamefont {C.~T.}\ \bibnamefont
  {Schmiegelow}}, \bibinfo {author} {\bibfnamefont {H.}~\bibnamefont
  {Kaufmann}}, \bibinfo {author} {\bibfnamefont {T.}~\bibnamefont {Ruster}},
  \bibinfo {author} {\bibfnamefont {J.}~\bibnamefont {Schulz}}, \bibinfo
  {author} {\bibfnamefont {V.}~\bibnamefont {Kaushal}}, \bibinfo {author}
  {\bibfnamefont {M.}~\bibnamefont {Hettrich}}, \bibinfo {author}
  {\bibfnamefont {F.}~\bibnamefont {Schmidt-Kaler}}, \ and\ \bibinfo {author}
  {\bibfnamefont {U.~G.}\ \bibnamefont {Poschinger}},\ }\href {\doibase
  10.1103/PhysRevLett.116.033002} {\bibfield  {journal} {\bibinfo  {journal}
  {Phys. Rev. Lett.}\ }\textbf {\bibinfo {volume} {116}},\ \bibinfo {pages}
  {033002} (\bibinfo {year} {2016})}\BibitemShut {NoStop}%
\bibitem [{\citenamefont {M{\o}lmer}\ and\ \citenamefont
  {S{\o}rensen}(1999)}]{molmer1999multiparticle}%
  \BibitemOpen
  \bibfield  {author} {\bibinfo {author} {\bibfnamefont {K.}~\bibnamefont
  {M{\o}lmer}}\ and\ \bibinfo {author} {\bibfnamefont {A.}~\bibnamefont
  {S{\o}rensen}},\ }\href@noop {} {\bibfield  {journal} {\bibinfo  {journal}
  {Physical Review Letters}\ }\textbf {\bibinfo {volume} {82}},\ \bibinfo
  {pages} {1835} (\bibinfo {year} {1999})}\BibitemShut {NoStop}%
\bibitem [{\citenamefont {Gilmore}\ \emph {et~al.}(2017)\citenamefont
  {Gilmore}, \citenamefont {Bohnet}, \citenamefont {Sawyer}, \citenamefont
  {Britton},\ and\ \citenamefont {Bollinger}}]{gilmore2017amplitude}%
  \BibitemOpen
  \bibfield  {author} {\bibinfo {author} {\bibfnamefont {K.~A.}\ \bibnamefont
  {Gilmore}}, \bibinfo {author} {\bibfnamefont {J.~G.}\ \bibnamefont {Bohnet}},
  \bibinfo {author} {\bibfnamefont {B.~C.}\ \bibnamefont {Sawyer}}, \bibinfo
  {author} {\bibfnamefont {J.~W.}\ \bibnamefont {Britton}}, \ and\ \bibinfo
  {author} {\bibfnamefont {J.~J.}\ \bibnamefont {Bollinger}},\ }\href@noop {}
  {\bibfield  {journal} {\bibinfo  {journal} {Physical review letters}\
  }\textbf {\bibinfo {volume} {118}},\ \bibinfo {pages} {263602} (\bibinfo
  {year} {2017})}\BibitemShut {NoStop}%
\bibitem [{\citenamefont {G.J.}\ \emph {et~al.}()\citenamefont {G.J.},
  \citenamefont {S.},\ and\ \citenamefont {D.F.V.}}]{milburn_phase_gate}%
  \BibitemOpen
  \bibfield  {author} {\bibinfo {author} {\bibfnamefont {M.}~\bibnamefont
  {G.J.}}, \bibinfo {author} {\bibfnamefont {S.}~\bibnamefont {S.}}, \ and\
  \bibinfo {author} {\bibfnamefont {J.}~\bibnamefont {D.F.V.}},\ }\href
  {\doibase 10.1002/1521-3978(200009)48:9/11<801::AID-PROP801>3.0.CO;2-1}
  {\bibfield  {journal} {\bibinfo  {journal} {Fortschritte der Physik}\
  }\textbf {\bibinfo {volume} {48}},\ \bibinfo {pages} {801}}\BibitemShut
  {NoStop}%
\bibitem [{\citenamefont {Dylewsky}\ \emph
  {et~al.}(2016{\natexlab{a}})\citenamefont {Dylewsky}, \citenamefont
  {Freericks}, \citenamefont {Wall}, \citenamefont {Rey},\ and\ \citenamefont
  {Foss-Feig}}]{DylewskyPRA}%
  \BibitemOpen
  \bibfield  {author} {\bibinfo {author} {\bibfnamefont {D.}~\bibnamefont
  {Dylewsky}}, \bibinfo {author} {\bibfnamefont {J.~K.}\ \bibnamefont
  {Freericks}}, \bibinfo {author} {\bibfnamefont {M.~L.}\ \bibnamefont {Wall}},
  \bibinfo {author} {\bibfnamefont {A.~M.}\ \bibnamefont {Rey}}, \ and\
  \bibinfo {author} {\bibfnamefont {M.}~\bibnamefont {Foss-Feig}},\ }\href
  {\doibase 10.1103/PhysRevA.93.013415} {\bibfield  {journal} {\bibinfo
  {journal} {Physical Review A}\ }\textbf {\bibinfo {volume} {93}},\ \bibinfo
  {pages} {1} (\bibinfo {year} {2016}{\natexlab{a}})},\ \Eprint
  {http://arxiv.org/abs/1510.05003} {arXiv:1510.05003} \BibitemShut {NoStop}%
\bibitem [{\citenamefont {Milburn}\ \emph {et~al.}(2000)\citenamefont
  {Milburn}, \citenamefont {Schneider},\ and\ \citenamefont
  {James}}]{milburn2000ion}%
  \BibitemOpen
  \bibfield  {author} {\bibinfo {author} {\bibfnamefont {G.}~\bibnamefont
  {Milburn}}, \bibinfo {author} {\bibfnamefont {S.}~\bibnamefont {Schneider}},
  \ and\ \bibinfo {author} {\bibfnamefont {D.}~\bibnamefont {James}},\
  }\href@noop {} {\bibfield  {journal} {\bibinfo  {journal} {Fortschritte der
  Physik}\ }\textbf {\bibinfo {volume} {48}},\ \bibinfo {pages} {801} (\bibinfo
  {year} {2000})}\BibitemShut {NoStop}%
\bibitem [{\citenamefont {Dylewsky}\ \emph
  {et~al.}(2016{\natexlab{b}})\citenamefont {Dylewsky}, \citenamefont
  {Freericks}, \citenamefont {Wall}, \citenamefont {Rey},\ and\ \citenamefont
  {Foss-Feig}}]{PhysRevA.93.013415}%
  \BibitemOpen
  \bibfield  {author} {\bibinfo {author} {\bibfnamefont {D.}~\bibnamefont
  {Dylewsky}}, \bibinfo {author} {\bibfnamefont {J.~K.}\ \bibnamefont
  {Freericks}}, \bibinfo {author} {\bibfnamefont {M.~L.}\ \bibnamefont {Wall}},
  \bibinfo {author} {\bibfnamefont {A.~M.}\ \bibnamefont {Rey}}, \ and\
  \bibinfo {author} {\bibfnamefont {M.}~\bibnamefont {Foss-Feig}},\ }\href
  {\doibase 10.1103/PhysRevA.93.013415} {\bibfield  {journal} {\bibinfo
  {journal} {Phys. Rev. A}\ }\textbf {\bibinfo {volume} {93}},\ \bibinfo
  {pages} {013415} (\bibinfo {year} {2016}{\natexlab{b}})}\BibitemShut
  {NoStop}%
\end{thebibliography}%

\begin{widetext}
\title{Supplementary Materials}
\maketitle

\newpage
\appendix\newpage\markboth{Appendix}{}
\setcounter{equation}{0}
\renewcommand{\thesection}{S}
\numberwithin{equation}{section}

\section{Supplementary Materials}
In this supplementary material we present supporting technical details for the main manuscript. In Sec.\ S1. we derive the total Hamiltonian with both a spin dependent force (SDF) and parametric amplification (PA), and we then consider the limitations of the rotating wave approximation (RWA) and the validity of the Lamb-Dicke regime in Sec.\ S2. In Sec.\ S3. we summarize the calculation of quantum spin squeezing in the presence of decoherence, and analyze the consequences of fluctuations in the system parameters.  In Sec.\ S4., we give further details on the fidelity of two-qubit gate using PA and its sensitivity to system parameter fluctuations.

\section{S1. Trapped ion Hamiltonian with SDF and PA\label{sec1}}
The Hamiltonian describing a crystal of $N$ trapped ions with two long-lived internal states can be written as
\begin{align}
\hat{\mathcal{H}}_{\rm ions}=\sum_{m=1}^N\omega_{m}\hat{a}^{\dagger}_{m}\hat{a}^{\phantom\dagger}_{m}+\frac{\omega_a}{2}\sum_{j=1}^{N}\hat{\sigma}_{j}^z,
\end{align}
where $\hat{a}^{\dagger}_{m}$ creates a collective excitation of the crystal with energy $\omega_m$, and $\omega_{a}$ is the qubit energy splitting. For simplicity we assume that the frequency $\omega_m$ decreases with the increasing mode number $m$. For the transverse modes of a linear ion string or a single plane crystal, the center-of-mass mode, equal to the single-particle trapping frequency (and referred to as the trap frequency), is the highest frequency mode $\omega_1$. There are many approaches to generating entanglement between trapped ions, but most of them share the common strategy of applying an oscillating spin-dependent dependent force.
This is often achieved using noncopropagating lasers to either drive stimulated Raman transitions \cite{molmer1999multiparticle} or to generate a spatially varying AC stark shift to the qubit transition \cite{milburn2000ion}, but recently it has also been achieved using strong magnetic field gradients in surface-electrode traps \cite{maggrad1,maggrad2}.  If the force has amplitude $F$, points in the $z$-direction, oscillates at a frequency $\mu$, and doesn't vary significantly over the length scale on which the ions motion is confined (for laser driven transitions, this is the so-called Lamb-Dicke regime), then the full Hamiltonian in the presence of the SDF is (in the rotating frame of the qubit and the oscillating force)
\begin{eqnarray}
\hat{\mathcal{H}}_{\text{SDF}}= -\hbar \sum_{m=1}^N\delta_m\hat{a}_m^{\dagger}\hat{a}_m+F\cos\left(\mu t\right)\sum_{i=1}^N\hat{z}_i(t)\hat{\sigma}_{i}^{z}.\label{eq:NHodf}
\end{eqnarray}
Here, $\delta_m=\mu-\omega_m$. Note that depending on the implementation, the force might couple to a Pauli matrix other than $\hat{\sigma}_i^z$. For simplicity here we assume the force couples to $\hat{\sigma}_i^z$, but note that our protocol for enhancing the spin-dependent force is directly applicable to the situation when a Pauli matrix other than $\hat{\sigma}_i^z$ is used (e.g. it works for the Molmer-Sorensen gate as well as the phase gate).  The ion position operators can be written $\hat{z}_i(t)=\sum_{m=1}^NU_{i,m}z_{0m}\big(e^{-i\mu t}\hat{a}_m+e^{i\mu t}\hat{a}_m^{\dagger}\big)$, where $z_{0m}=\sqrt{\hbar/2M\omega_m}$ and $U_{i,m}$ are the normal-mode transformation matrix elements obeying $\sum_{i=1}^NU_{i,m}U_{i,l}=\delta_{ml}$ and $\sum_{m=1}^NU_{i,m}U_{j,m}=\delta_{ij}$ \cite{james1998quantum}. Defining $f_m=F z_{0m}/2$ and rewriting the second term in terms of creation and annihilation operators, we obtain the Hamiltonian in Eq. (3) of the manuscript
\begin{align}
\hat{\mathcal{H}}_{\rm SDF}&=\hbar \sum_{m=1}^{N}\Big(f_m\big(\hat{a}_m+\hat{a}_m^{\dagger}\big)\sum_{i=1}^NU_{i,m}\hat{\sigma}_{i}^{z}-\delta_m\hat{a}^{\dagger}_m\hat{a}_m\Big)+\hat{\mathcal{H}}_{\rm CR},
\end{align}
with 
\begin{align}
\hat{\mathcal{H}}_{\rm CR}=\hbar \sum_{m=1}^{N}\Big(f_m\big(\hat{a}_me^{-2i\mu t}+\hat{a}_m^{\dagger}e^{2i\mu t}\big)\sum_{i=1}^NU_{i,m}\hat{\sigma}_{i}^{z}.
\end{align}

Parametric amplification can be achieved by modulating the voltage on the trap electrodes at a frequency close to twice that of a normal mode. Here we apply a voltage modulation $V(t)=-V\cos(2\mu t-\theta)$, which in the rotating frame of the SDF gives a contribution to the total Hamiltonian of \cite{Heinzen1990}
\begin{align}
\hat{\mathcal{H}}_{PA}&=\frac{2eV}{d^2_T}\cos(2\mu t-\theta)\sum_{i=1}^N\hat{z}_i(t)^2 =\cos(2\mu t-\theta)\sum_{m=1}^N \hbar g_m\left(\hat{a}_me^{i\mu t}+\hat{a}_m^{\dagger}e^{-i\mu t}\right)^2.
\end{align}
The final equality is obtained using $\sum_{i=1}^NU_{i,m}U_{i,l}=\delta_{ml}$, and we've defined $g_m\equiv eV/(M\omega_m d^2_T)$, where $e$ is the ion charge and $d_T$ is a characteristic trap dimension. Note that $g_m\approx g$ for all $m$ when the bandwidth of the normal modes is much smaller than the typical mode frequency.  The expression $g_1=eV/(M\omega_1 {d_T}^2)$ shows that large values of $g_1$ should be readily achievable with small $d_T$.  For the case of a quadratic trapping potential resulting from the application of a DC voltage $V_T$ to the trap electrodes, $d_T$ can be related to the center-of mass frequency $\omega_1$ as $(d_T)^{-2} \approx M(\omega_1)^2/(4eV_T)$ \cite{Heinzen1990}.  Straight forward algebra then results in the expression $g_1 \approx (V/V_T)\times (\omega_1/4)$.  Values of $g_1$ as large as $0.1\times \omega_1$ are feasible with a parametric drive $V$ comparable to the trap voltage $V_T$.

\

With both the SDF and PA (and ignoring the mode-dependence of $g_m$), the total Hamiltonian is
\begin{align}
\hat{\mathcal{H}}_{\rm T}=\hat{\mc{H}}_{\rm SDF} +\hat{\mc{H}}_{\rm PA} =\hbar \sum_{m=1}^{N}\Big[f_m\Big(\hat{a}_m+\hat{a}_m^{\dagger}\Big)\sum_{i=1}^NU_{i,m}\hat{\sigma}_{i}^{z}-\delta_m\hat{a}^{\dagger}_m\hat{a}_m+\frac{g}{2}\Big(\hat{a}^2_me^{-i\theta}+\hat{a}_m^{\dagger2}e^{i\theta}\Big)\Big]+\hat{\mathcal{H}}_{\rm CR},
\end{align}
where the counter-rotating terms of the total Hamiltonian are now given by
\begin{align}
\hat{\mathcal{H}}_{\rm CR}=\hbar \sum_{m=1}^{N}\Big[f_m\big(\hat{a}_me^{-2i\mu t}+\hat{a}_m^{\dagger}e^{2i\mu t}\big)\sum_{i=1}^NU_{i,m}\hat{\sigma}_{i}^{z}+g\cos\left(2\mu t-\theta\right)\hat{a}_m^{\dagger}\hat{a}_m+\frac{g}{2}\left(\hat{a}^2_me^{-i4\mu t+i\theta}+\hat{a}_m^{\dagger2}e^{i4\mu t-i\theta}\right)\Big].
\end{align}
When $|\delta_m|>|g|$, the time-independent part of $\hat{\mc{H}}_{\rm T}$ can be simplified using a Bogoliubov transformation $\hat{b}_m=\cosh r_m \hat{a}_m -e^{i\theta}\sinh r_m \hat{a}_m^{\dagger}$, with $e^{r_m}=[\left(\delta_m+g\right)/\left(\delta_m-g\right)]^{1/4}$ \cite{lemonde2016enhanced}, to obtain the total Hamiltonian 
\begin{align}
\hat{\mathcal{H}}_{\rm T}&=\hbar\sum_{m=1}^{N}\Big[ \Big(f^{\prime\ast}_m\hat{b}_m+f^{\prime}_m\hat{b}^{\dagger}_m\Big)\sum_{i=1}^N U_{i,m}\hat{\sigma}_{i}^{z}-\delta^{\prime}_m \hat{b}_m^{\dagger}\hat{b}_m\Big]+\hat{\mathcal{H}}_{\rm CR}. \label{eq:BmodeH}
\end{align}
Note that in terms of these transformed operators, the time-independent piece of $\hat{\mc{H}}_{\rm T}$ now takes the same form as the time-independent piece of $\hat{\mc{H}}_{\rm SDF}$, only with modified parameters $f^{\prime}_m= f_m(\cosh r_m +e^{i\theta}\sinh r_m)$ and $\delta^{\prime}_m=\sqrt{\delta_m^2-g^2}$. While the detunings $\delta^{\prime}_m$ of the Bogoliubov modes are independent of $\theta$, the degree of quadrature squeezing depends on $\theta$ as $\mathscr{S}_m=\left(\cosh 2r_m+\cos \theta \sinh 2r_m\right)^{-1/2}$.  The maximal quadrature squeezing $\mathscr{S}_m=e^{-r_m}$ occurs at $\theta=0$.

\section{S2. Validity of approximations\label{sec2}}

\subsection{S2.1. Rotating-wave approximation}

The dynamics induced by $\hat{\mathcal{H}}_{\rm T}$ are much simpler if the counter-rotating terms are dropped, which can in some situations be justified under the rotating wave approximation (RWA). To understand the validity of the RWA, we need to understand the effects due to $\hat{\mathcal{H}}_{\rm CR}$, which contains terms oscillating at frequencies of $2\mu$ and $4\mu$.  A quantitative assessment of the validity of RWA can be made via a brute-force numerical solution of the dynamics under the full Hamiltonian $\hat{\mathcal{H}}_{\rm T}$, but a qualitative appreciation for the importance of $\mathcal{H}_{\rm CR}$ can be gained by perturbative arguments.  In particular, writing $\hat{\mathcal{H}}_{\rm T}=\hat{\mathcal{H}}_{0}+\hat{\mathcal{V}}(t)$, with $\hat{\mc{V}}(t)$ being the counter-rotating Hamiltonian, we can readily calculate the time evolution operator in the interaction picture of $\hat{\mathcal{H}}_{0}$ to second order in time-dependent perturbation theory with respect to $\hat{\mathcal{V}}(t)$. To this order, we will find oscillating and secular ($\propto t$) contributions; the secular terms effectively shift the eigenvalues of $\mathcal{H}_0$, and a reasonable criterion for validity of the RWA is that these shifts do not appreciably affect the dynamics.

\

For simplicity we consider only the c.o.m.\ mode in what follows, in which case $\hat{\mathcal{H}}_0= \hbar f^{\prime}\big(\hat{b}+\hat{b}^{\dagger}\big)\big(\sum_{i=1}^N U_{i,1}\hat{\sigma}_{i}^{z}\big)-\hbar \delta^{\prime} \hat{b}^{\dagger}\hat{b}$ (generalization to the multi-mode case can be inferred at the end).  It will be convenient in what follows to separate the counter-rotating perturbation into pieces that oscillate at different frequencies as $\hat{\mathcal{V}}(t)=\hat{\mathcal{V}}^{2\mu}(t)+\hat{\mathcal{V}}^{4\mu}(t)$, with 
\begin{align}
\hat{\mathcal{V}}^{2\mu}(t)=\hbar f\Big(\hat{a}e^{-2i\mu t}+\hat{a}^{\dagger}e^{2i\mu t}\Big)\sum_{i=1}^NU_{i,1}\hat{\sigma}_{i}^{z}+\hbar g\cos\left(2\mu t-\theta\right)\hat{a}^{\dagger}\hat{a}, \,\,\,\,\,\,\,\,\,\,   \hat{\mathcal{V}}^{4\mu}(t)=\hbar\frac{g}{2}\left(\hat{a}^2e^{-i4\mu t}+\hat{a}^{\dagger2}e^{i4\mu t}\right).
\end{align}


The eigenstates of $\hat{\mathcal{H}}_0$ are product states of spin and motion.  For any spin state diagonal in the $z$ basis, $\ket{\sigma_1,\dots,\sigma_N}$, the corresponding motional eigenstate is easily obtained after the Bogoliubov transformation as
\begin{align}
\label{eq:eigenH0}
\ket{\tilde{n}}=\mathcal{D\left(\beta\right)}\ket{n},
\end{align}
where $D\left(\beta\right)=e^{\beta\hat{b}^{\dagger}-\beta^{\ast}\hat{b} }$, $\beta=(f^{\prime}/\delta^{\prime})\big(\sum_{i=1}^N U_{i,1}\sigma_{i}^{z}\big)$, and $\ket{n}$ is the Fock state \emph{of the transformed $b$-bosons}. The corresponding eigenenergies are $\mathcal{E}_n^{(0)}=-\hbar \delta^{\prime}\big(n-\beta^2\big)$.  The term proportional to $\beta^2$ is a spin-dependent shift of the zero-point energy, and is responsible for the induced spin-spin interactions. 

\

The solution of the Schr\"odingder equation $i\hbar \partial_t \ket{\psi(t)}=\big[\hat{\mathcal{H}}_0+\hat{\mathcal{V}}(t)\big]\ket{\psi(t)}$ can be written in the interaction picture of $\hat{\mathcal{H}}_0$ as $\ket{\psi(t)}=e^{-it\hat{\mathcal{H}}_0/\hbar}\ket{\psi_{I}(t)}$, where  $\ket{\psi_I(t)}=U_I(t)\ket{\psi(0)}$ and
\begin{align}
U_I(t)=1-\frac{i}{\hbar}\int_{0}^{t}dt_1 \hat{\mc{V}}_{I}(t_1)+\left(-\frac{i}{\hbar}\right)^2\int_{0}^{t}dt_1\int_{0}^{t_1}dt_2 \hat{\mc{V}}_{I}(t_1)\hat{\mc{V}}_{I}(t_2)+\cdots
\label{eq: perturbationU}
\end{align}
Here $\hat{\mc{V}}_{I}(t)=\hat{\mc{V}}_I^{2\mu}(t)+\hat{\mc{V}}_I^{4\mu}(t)$, with $\hat{\mc{V}}^{2\mu}_{I}(t)=e^{i t\hat{\mc{H}}_0/\hbar}\hat{\mathcal{V}}^{2\mu}(t)e^{- it\hat{\mc{H}}_0/\hbar}$ and $\hat{\mc{V}}^{4\mu}_{I}(t)=e^{i t\hat{\mc{H}}_0/\hbar}\hat{\mathcal{V}}^{4\mu}(t)e^{- it\hat{\mc{H}}_0/\hbar}$. The leading secular contributions to $U_I(t)$ come at second order in $\hat{\mc{V}}_{I}(t)$.  Since $V^{2\mu}_{I}(t)$ and $V^{4\mu}_{I}(t)$ have different oscillation frequencies, their cross terms in the expression $\hat{\mc{V}}_{I}(t_1)\hat{\mc{V}}_{I}(t_2)$ do not yield a secular contribution, and we can consider them independently. We consider first the contribution from $\hat{\mc{V}}_I^{4\mu}(t)$, which can be expanded as
\begin{align}
\hat{\mc{V}}^{4\mu}_{I}(t)&=\sum_{m,n}\ket{\tilde{m}}\bra{\tilde{m}}e^{\frac{i}{\hbar}\hat{\mathcal{H}}_0 t}\hat{\mc{V}}^{4\mu}(t)e^{-\frac{i}{\hbar}\hat{\mathcal{H}}_0t}\ket{\tilde{n}}\bra{\tilde{n}}=\hbar\frac{g}{2}\sum_{m,n}\bra{\tilde{m}}\hat{a}^{2}\ket{\tilde{n}}\ket{\tilde{m}}\bra{\tilde{n}}e^{-i\delta^{\prime}(m-n)t-4i\mu t}+\text{h.c.},
\end{align}
Using
\begin{align}
\bra{\tilde{m}}\hat{a}^{2}\ket{\tilde{n}}&=\left[e^{2r}\beta^2+(m+\frac{1}{2})\sinh 2r \right]\delta_{m,n}+2e^r\beta\sqrt{m+1}\cosh r \delta_{m,n-1}\nonumber\\
&+2e^r\beta\sqrt{m}\sinh r \delta_{m-1,n}+\sqrt{(m+1)(m+2)}\cosh^2r\delta_{m+1,n-1}+\sqrt{m(m-1)}\sinh^2r\delta_{m-1,n+1},
\end{align}
integrating over time, and using $\mu\gg \delta^{\prime}$, we find the secular diagonal contribution to $(-i/\hbar)^2\int_0^t dt_1\int_0^{t_1} dt_2 \hat{\mc{V}}^{4\mu}_{I}(t_1)\hat{\mc{V}}^{4\mu}_{I}(t_2)$ to be
\begin{align}
it\frac{g^2}{16\mu}\sum_{m,n}\left[|\bra{\tilde{m}}\hat{a}^{2}\ket{\tilde{n}}|^2-|\bra{\tilde{n}}\hat{a}^{2}\ket{\tilde{m}}|^2\right]\ket{\tilde{n}}\bra{\tilde{n}}=-\frac{it}{\hbar}\sum_{n}\mathcal{E}_n^{(2)}\ket{\tilde{n}}\bra{\tilde{n}},
\end{align}
where $\mathcal{E}_n^{(2)}=-\hbar \Delta (2\beta^2+n+1/2)$ and $\Delta\equiv g^2e^{2r}/(8\mu)=g^2/(8\mathscr{S}^2\mu)$. Similarly, we can find the energy shift due to $V^{2\mu}_I(t)$. Because $\bra{\tilde{m}}\hat{a}^{\dg}\hat{a}\ket{\tilde{n}}$ is symmetric about $m\leftrightarrow n$, the energy shift due to $V^{2\mu}_{I}(t)$ only comes from the term (proportional to $f$) that is linear in creation/anihilation operators.  After similar manipulations to above, we find a contribution on the order of $f^2/\mu$.  Because $f^2/\mu\ll \Delta (2\beta^2+n+1/2)$ for the parameters of interest, these corrections can be ignored relative to the energy shifts $\mc{E}^{(2)}_n$ computed above.
Summarizing the calculation so far, to second-order in perturbation theory we have energies given by 
\begin{align}
\mathcal{E}_n=-\hbar(\delta^{\prime}+\Delta)n+\hbar \beta^2 (\delta^{\prime}-2\Delta).
\end{align}
Comparing to the zeroth-order energies $\mathcal{E}_n^{(0)}=-\hbar \delta^{\prime}\big(n-\beta^2\big)$, we see that the perturbative corrections ($\propto\Delta$) increase the oscillation frequency (inferred from the coefficient of $n$) while reducing the spin-dependent geometric phase (inferred from the reduced coefficient of $\beta^2$). More generally, we expect the RWA to give a good estimate of the acquired geometric phase due to any particular mode whenever the correction to the coefficient of $\beta^2$ due to that mode, $\delta^{\prime}_{m}\rightarrow\delta^{\prime}_m-\Delta\delta^{\prime}_{m}$, with $\Delta\delta^{\prime}_m\equiv g^2/(4\mathscr{S}_m^2\mu)$, is insignificant.  We therefore require $\Delta\delta^{\prime}_m\ll \delta^{\prime}_m$.  As shown in Fig. 4 of the manuscript, numerical simulations including the CRW Hamiltonian start to deviate appreciably from the RWA results when $\Delta\delta^{\prime}_m=\delta^{\prime}_m/2$ using a trap frequency of $3.045$ MHz (the same trap frequency used for the calculation of Fig. 3).

\subsection{S2.2. Lamb-Dicke approximation}

The manuscript begins with the implicit assumption (in writing down $\hat{\mathcal{H}}_{\rm SDF}$) that the applied force is spatially uniform.  When the SDF is implemented via optical dipole forces, this assumption is valid in the so-called Lamb-Dicke regime.  Because the extent of this regime depends in detail on the motional states of the ions, which are modified by PA, we will briefly consider the validity of the Lamb-Dicke approximation in the presence of PA.  If the SDF is generated by noncopropagating lasers with wave-vector difference $\Delta k$, the Lamb-Dicke limit requires $\Delta k\langle\hat{z}_i^{2}\rangle^{1/2} \ll 1$, where  $\braket{\hat{z}_i^2}=\sum_m(U_{i,m})^2z^2_{0m}(2n_m+1)$ and $n_m$ is the time-averaged expectation value of $\hat{a}^{\dagger}_m\hat{a}_m$. For $n_1\gg1$, the contribution of the c.o.m.\ mode to $\Delta k^2\langle z_{i}^2\rangle$ is given approximately by $2\eta_1^2 n_1/N$, where $\eta_m\equiv \Delta k z_{0m}$ is the Lamb-Dicke parameter for the $m$th mode.  The value of $n_1$ is related to the geometric phase $\Phi$ and the spin states, and can be evaluated using Eq. \eqref{eq:eigenH0} as  $n_1\approx 3 \Phi\braket{S_z^2}/(\pi N\mathscr{S}^2)$. Therefore, the Lamb-Dicke limit requires (just considering the c.o.m.\ mode)
\begin{align}
\label{eq:lamb_dicke_constraint}
\mathscr{S}\gg \frac{\eta_1}{N}\sqrt{\frac{6 \Phi \braket{S_z^2}}{\pi}}.
\end{align}
The Lamb-Dicke parameter $\eta_1$ depends on the experiment, but is typically less than 0.2 \cite{Bohnet2016,Ballance:2016,PhysRevLett.120.020501}. For tasks such as spin squeezing $\Phi \sim N^{1/3} $ \cite{ma2011quantum} and $\braket{S_z^2}\sim N$,  in which case \eref{eq:lamb_dicke_constraint} becomes (ignoring order unity prefactors) $\mathscr{S} \gg \eta_1N^{-1/3}$.  Since the right hand side can be very small for large $N$, large enhancements ($\mathscr{S}\ll 1$) are consistent with the Lamb-Dicke limit.  For a two-qubit gate $\Phi=N\pi/4$  \cite{PhysRevLett.112.190502} and $\braket{S_z^2}\sim 1$.  In this case \eref{eq:lamb_dicke_constraint} becomes $\mathscr{S} \gg\eta_{1}N^{-1/2}$, again ensures that large enhancements are consistent with the Lamb-Dicke limit for large $N$. 

\section{S3. Quantum Spin Squeezing \label{sec3}}
Here we define the Ramsey squeezing parameter discussed in the main text, explain its saturation (as $N\rightarrow\infty$) in the presence of decoherence, and quantify its sensitivity to various experimental imperfections.

\subsection{S3.1. Squeezing with decoherence}
Quantum spin squeezing (QSS) along a direction rotated by $\psi$ (from $+z$ direction) about the $x$-axis is defined by
$\xi_{\psi}^2=N\Delta \hat{S}_{\psi}^2/|\braket{\bm{S}}|^2$,
where $\hat{S}_{\psi}=\cos(\psi)\hat{S}_z-\sin(\psi)\hat{S}_y$, $\Delta \hat{S}_{\psi}^2=\braket{ \hat{S}_{\psi}^2}- \braket{\hat{S}_{\psi}}^2$, 
and $\bm{S}=\frac{1}{2}\sum_i\big(\hat{\sigma}^x_{i},\hat{\sigma}^y_{i},\hat{\sigma}^z_{i}\big)$. The Ramsey spin squeezing is obtained by minimizing $\Delta \hat{S}_{\psi}^2$ over the angle $\psi$,
\begin{align}
\xi_R^2=\frac{N \min_{\psi}(\Delta \hat{S}_{\psi}^2)}{|\braket{\bm{S}}|^2}=\frac{N}{2|\braket{\bm{S}}|^2}\left(\Delta \hat{S}_{y}^2+\Delta \hat{S}_{z}^2-\sqrt{\big(\Delta \hat{S}_{y}^2-\Delta \hat{S}_{z}^2\big)^2+4\text{Cov}\big( \hat{S}_{y},  \hat{S}_{z}\big)^2}\right),
\label{eq:SSangle}
\end{align}
where the optimal angle is $\psi_{\text{opt}}=1/2\arctan\left[2\text{Cov}\big( \hat{S}_{y},  \hat{S}_{z}\big)/\big(\Delta \hat{S}_{y}^2-\Delta \hat{S}_{z}^2\big)\right]$ with $\text{Cov}\big( \hat{S}_{y},  \hat{S}_{z}\big)\equiv \braket{\hat{S}_{y}\hat{S}_{z}+\hat{S}_{z}\hat{S}_{y}}/2-\braket{\hat{S}_{y}}\braket{\hat{S}_{z}}$. In the presence of dephasing at rate $\Gamma_{\rm el}$ and spontaneuous spin flips from $\ket{\uparrow}$ to $\ket{\downarrow}$ ($\ket{\downarrow}$ to $\ket{\uparrow}$) at rate $\Gamma_{\rm ud}$ ($\Gamma_{\rm du}$), spin dynamics due to an Ising interaction in the $\hat{z}$ basis can be modeled by a master equation in the Lindblad form, which can be solved exactly \cite{foss2013nonequilibrium}. For completeness, we quote the spin correlation functions in the case of uniform coupling, i.e. $J_{ij}=J$ for all $i\ne j$,
\begin{align}
\label{eq:SScorrelation}
\braket{\hat{\sigma}^{+}_i}=\frac{e^{-\Gamma t}}{2}\Phi^{N-1}(J,t), \,\,\,\,\,\,\,\,\,\,\, \braket{\hat{\sigma}_i^a\hat{\sigma}_j^b}=\frac{e^{-2\Gamma t}}{4}\Phi^{N-2}((a+b)J,t), \,\,\,\,\,\,\,\,\,\,\, \braket{\hat{\sigma}_i^a\hat{\sigma}_j^z}=\frac{e^{-\Gamma t}}{2}\Psi(aJ,t)\Phi^{N-2}(aJ,t),
\end{align} 
where $a,b\in \{+,-\}$ and 
\begin{align}
\Phi(J,t)&=e^{-\frac{\left(\Gamma_{\text{ud}}+\Gamma_{\text{du}}\right)t}{2}}\left[\cos\left(t\sqrt{(2i\gamma+2J/N)^2-\Gamma_{\text{ud}}\Gamma_{\text{du}}}\right)+t\frac{\Gamma_{\text{ud}}+\Gamma_{\text{du}}}{2}\text{sinc}\left(t\sqrt{(2i\gamma+2J/N)^2-\Gamma_{\text{ud}}\Gamma_{\text{du}}}\right)\right],\\
\Psi(J,t)&=e^{-\frac{\left(\Gamma_{\text{ud}}+\Gamma_{\text{du}}\right)t}{2}}t\left[i(2i\gamma+2J/N)-2\gamma\right]\text{sinc}\left(t\sqrt{(2i\gamma+2J/N)^2-\Gamma_{\text{ud}}\Gamma_{\text{du}}}\right).
\end{align}
The above expressions are obtained assuming the initial state is a product state with all spins pointed along the $x$ direction, and also assuming that the final state has no spin-motion entanglement. Here $\gamma=\left(\Gamma_{\text{ud}}-\Gamma_{\text{du}}\right)/4$, $\Gamma=(\Gamma_{\text{r}}+\Gamma_{\text{el}})/2$, and $\Gamma_{\text{r}}=\Gamma_{\text{ud}}+\Gamma_{\text{du}}$. To calculate spin squeezing, we also need the expressions for $\braket{\hat{\sigma}_i^z}$ and $\braket{\hat{\sigma}_i^z\hat{\sigma}_j^z}$. Using the master equation in Ref. \cite{foss2013nonequilibrium}, we find 
\begin{align}
\braket{\hat{\sigma}_i^z}=\frac{4\gamma}{\Gamma_r}\left(e^{-\frac{1}{2}\Gamma_rt}-1\right), \,\,\,\,\,\,\,\, \braket{\hat{\sigma}_i^z\hat{\sigma}_j^z}=\braket{\hat{\sigma}_i^z}\braket{\hat{\sigma}_j^z}.
\end{align}
For the parameters of interest, e.g. $\gamma\ll\Gamma_r$ and $\Gamma_r t\ll1$, $\braket{\hat{\sigma}_i^z}\approx 0$ and $\braket{\hat{\sigma}_i^z\hat{\sigma}_j^z}\approx 0$. Therefore, the contrast of the total spins is
\begin{align}
\label{eq:contrast}
|\braket{\bm{S}}|\approx\frac{N}{2}\sqrt{\braket{\hat{\sigma}^{x}_i}^2+\braket{\hat{\sigma}^{y}_i}^2}=N|\braket{\hat{\sigma}^{+}_i}|=\frac{N}{2}e^{-\Gamma t}\Phi^{N-1}(J,t),
\end{align}
where the overall exponential decay is due to the spontaneous photon scattering and the $\Phi^{N-1}(J,t)$ term captures the interplay between the Raman decoherence and many-body dynamics \cite{Bohnet2016, foss2013nonequilibrium}. For the calculation of Fig. 2, we use Eqs. \eqref{eq:SScorrelation} through \eqref{eq:contrast} along with parameters typical of that used in \cite{Bohnet2016}, in particular $\gamma=0$, $\Gamma_{\text{el}}=0.12 J$, and $\Gamma_{\text{r}}=0.04 J$ to calculate the Ramsey squeezing parameter given by Eq. \eqref{eq:SSangle} and minimize $\xi_R^2$ with respect to time. 

To understand the saturation of $\xi_R^2$ with $N$ due to Raman scattering, we consider the relatively simpler (but nevertheless representative) case when $\Gamma_{\text{ud}}=\Gamma_{\text{du}}$.  In the limit $N\gg J/\Gamma_{\text{r}}$ we can expand $\Phi(J,t)$ and $\Psi(J,t)$ to the leading order in $Jt$ and $\Gamma_{\rm r} t$ to arrive at
\begin{align}
\xi_R^2\approx 1+A-\sqrt{A^2+B^2},
\end{align}
where $A=2(Jt)^2$ and $B=2Jt-\Gamma_{\rm r}Jt^2$. Here we assumed $e^{-\Gamma t}\approx1$ and  $\Phi^{N-1}(J,t)\approx 1$ under the limits $\Gamma \sim\Gamma_{\text{r}}\ll J$ and $1\ll J/\Gamma_{\text{r}}\ll N$, respectively. These approximations can be justified by substituting the optimal time derived below into the expressions $e^{-\Gamma t}$ and $\Phi^{N-1}(J,t)$. In this way, we find the minimum value of $\xi_R^2\approx 3\left(\Gamma_{\rm r}/4J\right)^{2/3} $ at the optimal time $t_{\text{opt}}=(J/2\Gamma_{\rm r})^{1/3}/J$. 
The same scaling has been obtained using a different method \cite{PhysRevLett.121.070403}, in which the spin squeezing is obtained from decoherence-free and decoherence-only contributions, separately, before optimizing over the interaction time.

\subsection{S3.2. Technical limitations}

The spin squeezing parameter $\xi_R^2$ as given by \eref{eq:SSangle} can be optimized with respect to the interaction time $t$ for fixed $\theta$, $\delta$, $g$ and $\Gamma$. 
However, fluctuations of system parameters can lead to reductions of the achieved spin squeezing relative to its optimal value.  This statement is true with or without PA, but PA introduces some qualitatively new effects that warrant attention: (1) Fluctuations of the phase $\theta$ (which plays no role without PA) modify the accumulated geometric phase, and (2) Fluctuations in $\delta$ or $g$, or $t$ modify the accumulated geometric phase \emph{and} can prevent the c.o.m.\ phase-space trajectory from closing (both of these effects could be exaggerated by PA).

\subsubsection{Fluctuations in $\theta$}
Especially for SDFs implemented by optical dipole forces, phase locking of the SDF and PA  is likely to be technically demanding, warranting a careful analysis of the effects of fluctuations in their relative phase $\theta$. As discussed in the manuscript, the period of the c.o.m.\ phase-space trajectory is independent of $\theta$, and therefore fluctuations in $\theta$ do not cause residual spin-motion entanglement. However, the geometric phase enclosed by the c.o.m.\ trajectory \emph{does} depend on $\theta$, implying that fluctuations in $\theta$ will result in fluctuations of the effective spin-spin interaction strength $J$.  We will show that $\xi_R^2$ depends only quartically on $\theta$ for small $\theta$, implying that even relatively poor experimental control over the relative phase between the SDF and PA can be tolerated.

\

When $J$ fluctuates around its optimal value, $J\rightarrow J+\Delta J$, the spin squeezing parameter $\xi_R^2$ can be expanded in $\Delta J$ (from here onwards, the symbol $\Delta$ denotes a small deviation from the optimal value of a parameter) as
\begin{align}
\xi_R^2(J+\Delta J)\approx \xi_R^2(J)+\frac{1}{2}\frac{d^2\xi_R^2(J)}{dJ^2}\Delta J^2 \approx \xi_R^2(J)+\frac{\Delta J^2}{J^2},
\label{eq:xifluct}
\end{align}
where the final approximation holds in the presence of decoherence as long as  $N\ll J/\Gamma_{r}$ (i.e.\ when the spin-squeezing has not yet saturated).  The optimal (maximum) value of the spin-spin interaction is $J=f^2/(\delta-g)$, occurring when $\theta=0$. The explicit dependence of $J+\Delta J$ on $\theta$ is given by
\begin{align}
\label{eq:J_of_theta}
J+\Delta J=\frac{f^{\prime2}}{\delta^{\prime}}=\frac{f^2}{\delta-g}\frac{1+\cos\theta}{2}+\frac{f^2}{\delta+g}\frac{1-\cos\theta}{2}\approx \frac{f^2}{\delta-g}\bigg(1-\frac{\theta^2}{4}\bigg),
\end{align}
where in the last line we have assumed that $\delta-g\ll\delta+g$ and that $\theta$ is small. Equation \eqref{eq:J_of_theta} implies that $\Delta J\approx -J\theta^2/4$, which can be inserted into \eref{eq:xifluct} to yield 
\begin{align}
\Delta\xi^2_{\rm R}=\xi^2_{\rm R}(J+\Delta J)-\xi^2_{\rm R}(J)\approx \frac{\theta^4}{16},
\end{align}
showing that the spin squeezing is only quartically dependent on the phase $\theta$.

\subsubsection{Fluctuations in $\delta$}

Fluctuations in $\delta$, $g$, or $t$ all have a qualitatively similar effect, in that they modify both the aqcuired geometric phase and result in residual spin-motion entanglement.  Fluctuations in $\delta$, caused by underlying fluctuations in either the motional mode frequencies or the SDF frequency $\mu$, are likely to place the most severe limitation on spin squeezing, and we therefore only consider fluctuations in $\delta$ explicitly here.  A naive procedure to account for fluctuations in $\delta$ would be to simply use the correlation functions given in Sec.\ \ref{sec3} and propagate the fluctuations in $\delta$ through to fluctuations in $J$.  Strictly speaking this is not justified, because when $\delta$ is not precisely tuned to disentangle the c.o.m\ mode at the desired time $t$, these expressions are no longer correct. In the absence of decoherence, spin-squeezing can be computed exactly in the presence of spin-motion entanglement  \cite{PhysRevA.93.013415}; a careful analysis of the consequences of spin-motion entanglement shows that for large $N$, this effect is insignificant compared to the modification due to the naive analysis proposed above.  Therefore, we can justifiably proceed by treating fluctuations in $\delta$ by inferring the induced fluctuations in $J$ in the expression for $\xi^2_{\rm R}$ given in Sec.\ \ref{sec3}.  With PA, fluctuations in $\delta$ propagate to $J=f^2/\left(\delta-g\right)$ (assuming $\theta=0$) as
\begin{align}
\frac{\Delta J}{J} = -\frac{\Delta \delta}{\delta-g} \approx-\frac{\Delta \delta}{\delta}\frac{1}{2\mathscr{S}^{4}}.
\end{align}
Plugging this result into \eref{eq:xifluct}, we see that spin squeezing depends quadratically on $\Delta \delta/\delta$ (as it would at $g=0$), but with a prefactor that grows (through the dependence on $\mathscr{S}$) as the enhancement increases.

\section{S4. Fidelity of a Two-Qubit Gate \label{sec4}}
\subsection{S4.1. Optimal Fidelity}
Even assuming perfect experimental control, a two-qubit gate in a multi-ion system still incurs an error due to residual spin-motion entanglement at the gate time, which can be quantified by  by \cite{PhysRevLett.120.020501}
\begin{align}
\epsilon_0(\delta,g,t)=\sum_{m=1}^N \left|\alpha_m(\delta_m^{\prime},t)\right|^2,
\label{eq:infidelity}
\end{align}
where $\alpha_m(\delta_m^{\prime},t)=f_m/\delta_m^{\prime}\left[\sin (\delta_m^{\prime}t)+i(\cos(\delta_m^{\prime}t)-1)/\mathscr{S}_m^2\right]$ is the residual displacement of the $m$th mode at the time $t$ and $\delta_m^{\prime}=\sqrt{\delta_m^2-g^2}$. In the above equation, the dependence of $\epsilon_0(\delta,g,t)$ on $\delta=\mu-\omega_1$ is obtained by assuming stable mechanical frequencies $\omega_m$ such that $\delta_m$ only depends on $\mu$ (and implicitly on $\delta$). For fixed $\delta$, $g$, and $f$, the fidelity will be optimized at some time $t=t_{\rm opt}$, yielding the optimal fidelity $F(\delta,g,t_{\rm opt})=1-\epsilon_0(\delta,g,t_{\rm opt})$. Since the c.o.m. mode is amplified the most significantly among all the modes, the optimal time satisfies $\delta^{\prime}t_{\rm opt} \approx 2\pi$. For $m\ne1$, the maximum residual displacement of each mode is
\begin{align}
|\alpha_m(\delta_m^{\prime},\pi/\delta_m^{\prime})|=\frac{2f_m}{\delta_m^{\prime}\mathscr{S}_m^2}=\frac{2f_m}{\delta_m-g}<\frac{2f_m}{\omega_1-\omega_m},
\end{align}
where we have used the condition $\delta-g>0$ to get the inequality. Therefore, the maximum residual displacement is independent of the squeezing factor $\mathscr{S}$ or the parametric driving strength $g$.

  In the manuscript, we use Eq. \eqref{eq:infidelity} to plot the fidelity of a two-qubit gate between two neighboring ions on either end of a $5-$ion chain as a function of $g$ (blue dotted curve in Fig.3 of the manuscript) for a gate time $t_{\rm opt}$ between $190 \ \mu$s and $160 \ \mu$s for conditions of a recent experiment using a modulated pulsed laser \cite{PhysRevLett.120.020501}.  A trap frequency of $3.045$ MHz (same as Fig. 4) is assumed. Note that we fix the parameter $\tau=2\pi/\sqrt{\delta^2-g^2}=180 \ \mu$s, however the actual optimal gate time varies for two reasons: 1) the effective period is smaller than $\tau$ due to the counter-rotating terms; 2) $t_{\rm opt}$ is optimized around this effective period.

\subsection{S4.2. Technical limitations}
The maximum fidelity derived above will be reduced by technical fluctuations in the parameters $\delta$, $g$, and $t$. As above, we treat fluctuations in $\delta$ as a representative example.  The analysis can be directly applied to the other parameters, although we expect their fluctuations to be relatively less important in typical experiments. The error due to fluctuations in $\delta$ around its optimal value can be broken into two qualitatively different pieces $\epsilon_0$ and $\epsilon_1$. First, the residual displacements of all the modes $\epsilon_0(\delta,g,t_{\rm opt})$ will be modified. We can expand $\epsilon_0(\delta+\Delta \delta,g,t_{\rm opt})$ to give the leading contribution from modifications of the c.o.m. mode trajectory (which plays the most significant role) as
\begin{align}
\epsilon_0(\delta+\Delta \delta,g,t_{\rm opt})\approx \epsilon_0(\delta,g,t_{\rm opt})+\left(\frac{f}{\delta^{\prime}}\right)^2\left(\frac{\delta \Delta\delta}{\delta^{\prime }}t_{\rm opt}\right)^2\approx \epsilon_0(\delta,g,t_{\rm opt})+\pi^2\left(\frac{f}{\delta}\right)^2\left(\frac{\Delta\delta}{\delta}\frac{1}{4\mathscr{S}^6}\right)^2,
\end{align}
where we have used $\delta^{\prime}t_{\rm opt} \approx 2\pi$ and $\delta-g\ll \delta+ g$ to get the second approximation. For a fair comparison with the standard trapped ions without the parametric amplification, we consider a fixed gate time $t_{\rm opt}$ and a fixed geometric area. According to Table I in the main text, $(f/\delta)^2\propto \mathscr{S}^6$ and $1/\delta\propto \mathscr{S}^2$. Overall, the frequency-fluctuation error is proportional to $1/\mathscr{S}^2$.

Second, the geometric phase for a two-qubit gate will fluctuate around its optimal value $N\pi/4$ \cite{PhysRevLett.112.190502} through the relation $\Phi =2 f^2t_{\rm opt}/(\delta-g)$. This fluctuation  $\Delta\Phi$ will reduce the fidelity by
\begin{align}
\epsilon_1(\delta+\Delta\delta,g,t_{\rm opt})\approx\left(\frac{\Delta\Phi}{N}\right)^2\approx\left(\frac{\pi}{4}\right)^2\left(\frac{\Delta\delta}{\delta}\frac{1}{2\mathscr{S}^4}\right)^2.
\end{align}

Therefore, the shift of the fidelity due to $\Delta\delta$ can be estimated as
\begin{align}
\Delta F=F(\delta+\Delta\delta,g,t_{\rm opt})- F(\delta,g,t_{\rm opt})\approx-\left(\frac{\pi}{4}\right)^2\left[1+\left(\frac{2f}{\delta\mathscr{S}^2}\right)^2\right]\left(\frac{1}{2\mathscr{S}^4}\right)^2\left(\frac{\Delta\delta}{\delta}\right)^2.
\end{align}

Assuming $\Delta\delta$ is a Gaussian random variable with a zero mean and the standard deviation $\sigma_{\delta}$, such that $\braket{(\Delta \delta)^2}=\sigma_{\delta}^2$ and $\braket{(\Delta \delta)^4}=3\sigma_{\delta}^4$, we obtain the standard deviation of the fidelity shift due to $\Delta\delta$ as
\begin{align}
\sqrt{\braket{\Delta F^2}-\braket{\Delta F}^2}=\sqrt{2}\left(\frac{\pi}{4}\right)^2\left[1+\left(\frac{2f}{\delta\mathscr{S}^2}\right)^2\right]\left(\frac{1}{2\mathscr{S}^4}\right)^2\left(\frac{\sigma_{\delta}}{\delta}\right)^2 .\end{align}
For the simulation in Fig.\ (3) of the main manuscript, we have chosen $f/2\pi\approx 1.2$ kHz, $\delta/2\pi \approx 35$ kHz, $\delta^{\prime}/2\pi=1/\tau$ and $\tau\approx 0.165$ ms. For a standard deviation $\sigma_{\delta} = 0.21$ kHz we find $\sqrt{\braket{\Delta F^2}-\braket{\Delta F}^2}\approx 0.6\%$, giving a total fidelity that remains above $99\%$.

\end{widetext}

\end{document}